\documentclass[12pt]{article}

\usepackage[margin=2.5cm]{geometry}

\usepackage[utf8]{inputenc}
\usepackage[math]{easyeqn}
\usepackage{graphicx}
\usepackage{stackrel,amssymb}
\usepackage{stackengine,scalerel}
\usepackage{url}
\usepackage{listings}
\usepackage{todonotes}

\newcommand{\PE}{{PE}}
\newcommand{\LB}{{LB}}
\newcommand{\CE}{{CE}}

\newcommand{\lrarrow}{\mathrel{\ensurestackMath{\ThisStyle{%
  \stackanchor[\dimexpr-2.5pt-4\LMpt]{\SavedStyle\,\,\longrightarrow}%
                                     {\SavedStyle\longleftarrow\,\,}}}}}
                                     
\begin{document}

\title   {Dynamic resource allocation for efficient parallel CFD simulations}
\author{G. Houzeaux, R.M. Badia, R. Borrell, D. Dosimont, \\ J. Ejarque, M. Garcia-Gasulla, V. L\'opez \\                 
Barcelona Supercomputing Center, Torre Girona, \\ c/ Jordi Girona 31, 08004 Barcelona, Spain}

\date{}

\maketitle

 \begin{abstract}
CFD users of supercomputers usually resort to rule-of-thumb methods to select the number of subdomains (partitions) when relying on MPI-based parallelization. One common approach is to set a minimum number of elements or cells per subdomain, under which the parallel efficiency of the code is "known" to fall below a subjective level, say 80$\%$. The situation is even worse when the user is not aware of the “good” practices for the given code and a huge amount of resources can thus be wasted. This work presents an elastic computing methodology to adapt at runtime the resources allocated to a simulation automatically. The criterion to control the required resources is based on a runtime measure of the communication efficiency of the execution. According to some analytical estimates, the resources are then expanded or reduced to fulfil this criterion and eventually execute an efficient simulation.
\end{abstract}



\section{Introduction}

\begin{figure}
  \centering
  \includegraphics[width=0.49\textwidth]{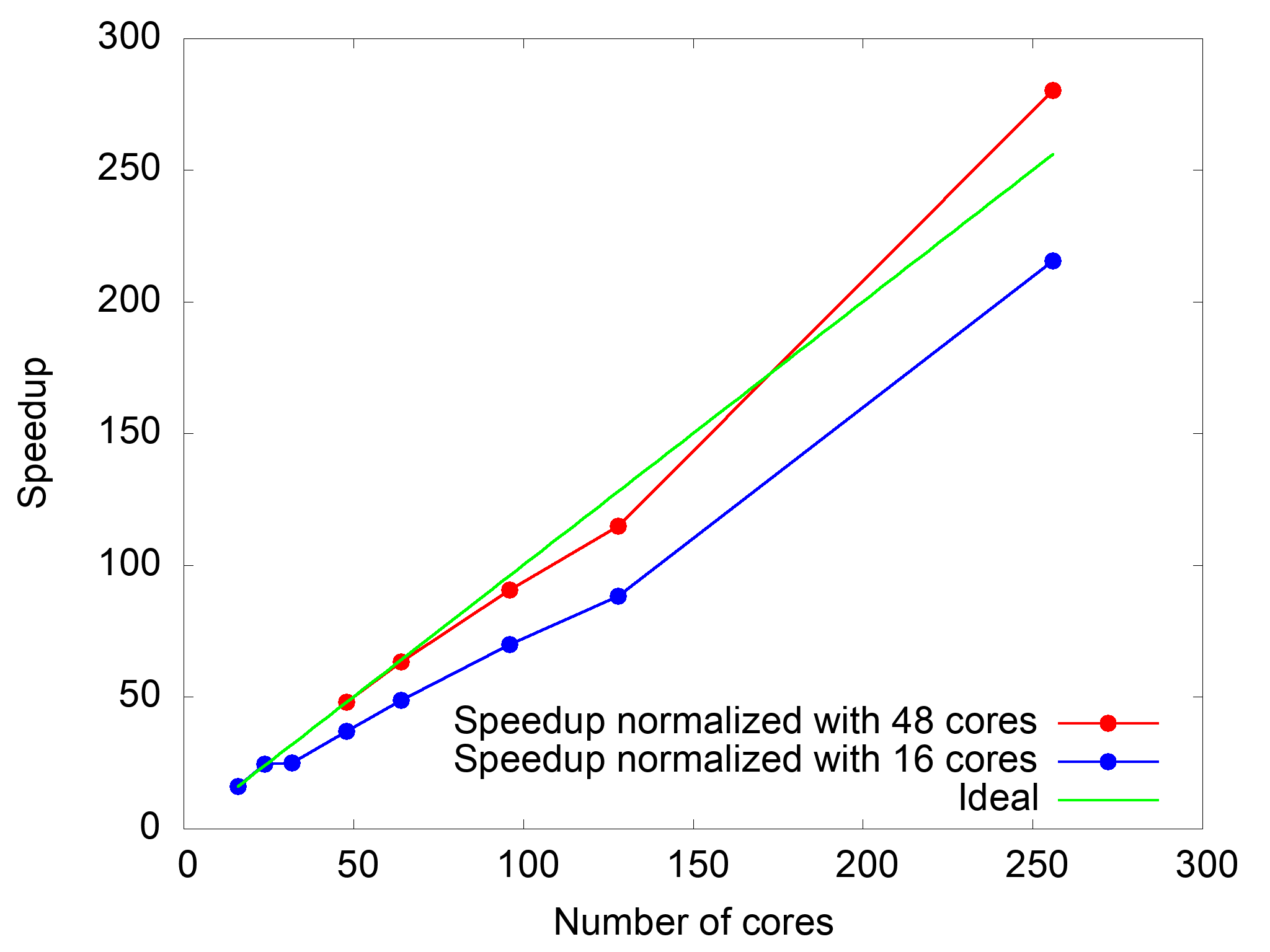}
  \includegraphics[width=0.49\textwidth]{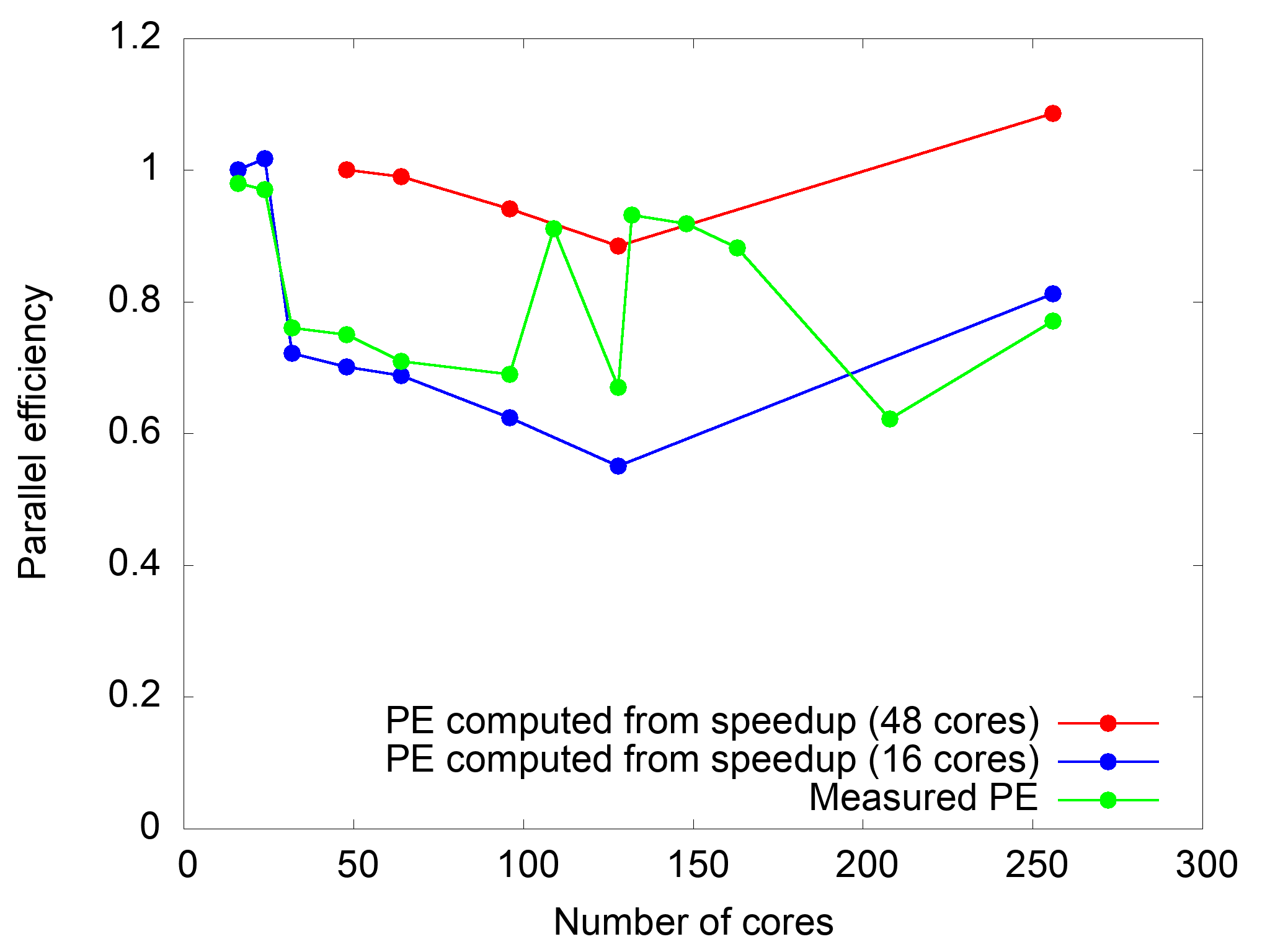}
  \caption{Speedup and parallel efficiency. 
  (Left) Relative speedup measure normalized with the CPU time obtained on 16 and 48 cores.
  (Right) Relative parallel efficiency ($\PE$) computed from speedup compared to absolute parallel efficiency.}
  \label{fig:speedup-generic}
\end{figure}

Computational Fluid Dynamics (CFD) is probably the field of computational continuum mechanics that traditionally consumes most of the worldwide available computational resources. The majority of CFD codes rely on substructuring techniques, and communications are habitually handled by the MPI library \cite{mpi3}. When running CFD simulations, the users usually resort to rule-of-thumb methods to select the number of subdomains of the partition. One common approach is to set a minimum number of elements/nodes/cells per subdomain, under which the “parallel” efficiency of the code is "known" to fall below a target level (say 80\%), usually set by the operation department of the supercomputing centers, but hardly controlled. The situation is even worse when the user is not aware of the best practices for a given code and a huge amount of resources can thus be poorly used. 
In addition, should the parallel efficiency of the code be known to the user, its value is usually relative to a base run as it is computed from speedups normalized at a high core count and for generic test cases, which can result in completely erroneous values \cite{ueabs2}. Figure \ref{fig:speedup-generic} illustrates this issue. 
On the left figure, two speedups of the same CFD simulation are shown, normalized with the computation time obtained with 16 cores and 48 cores. We observe that even if the tendency is similar, a huge difference arises when considering 256 cores. As common practice, the parallel efficiency is computed from the actual relative speedup and the ideal one. In this particular case, we thus have a parallel efficiency greater than 1 when normalizing with 48 cores. The right figure shows the real measured parallel efficiency, by the library TALP (Tracking Application Low-level Performance) \cite{talp} that will be introduced in Section \ref{sec:talp}. We observe a large discrepancy between normalized results and TALP measurements.
For example, when considering 256 cores, the difference is 20\% between the one normalized using 48 cores and the measured one.\\

From runtime performance measurements, to enable efficient CFD simulations, we propose a methodology to adjust the computing resources allocated for the simulation automatically.
More specifically, we concentrate on controlling the communication efficiency, which is, together with the load balance, the driving performance metric of parallel efficiency. Following this measure, resources are expanded or reduced to fulfill the target criterion which eventually ensures efficient utilization of the computing resources. 
This mechanism that adapts the resources automatically is usually referred to as elasticity in the computer science jargon. In addition, this elastic computing proposed is achieved remaining within the same SLURM job \cite{slurm}: this makes the proposed strategy automatic and transparent to the user.\\

\begin{figure}
  \centering
  \includegraphics[width=0.99\textwidth]{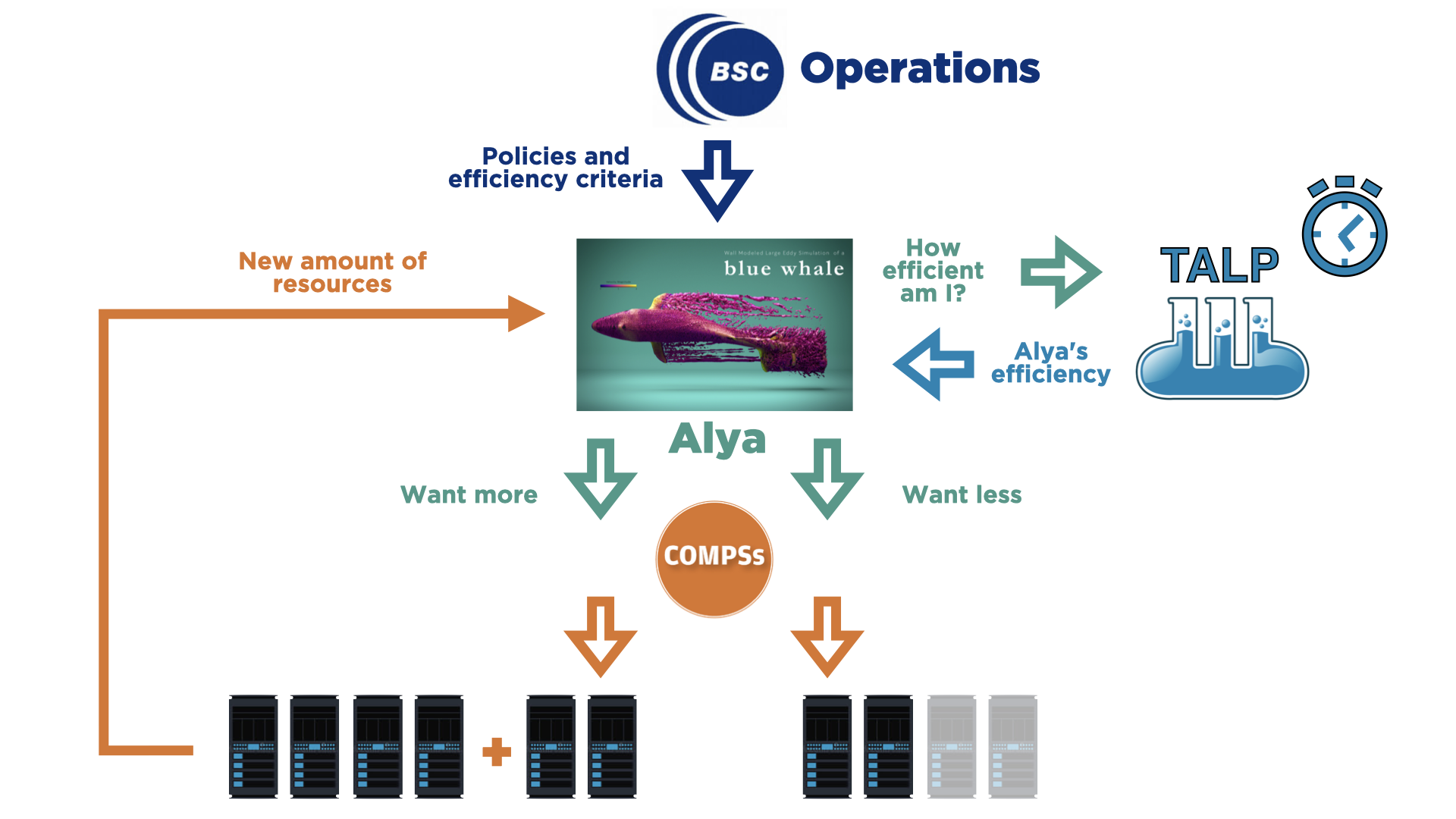}
  \caption{Optimizing the resources. Workflow for elastic computing of CFD simulations, involving different codes and libraries: Alya (CFD), TALP (efficiency measures) and COMPSs (elastic computing).}
  \label{fig:workflow}
\end{figure}

The optimization workflow, including its actors, is illustrated in Figure \ref{fig:workflow}. Alya is the CFD code, described in Section \ref{sec:alya}. TALP is the library linked to Alya to measure the communication efficiency used as a target criterion and is described in Section \ref{sec:talp}. Eventually, COMPSs is the library in charge of controlling the workflow and interacting with the batch system SLURM. It is described in Section \ref{sec:compss}.
As a control parameter of the workflow, the user prescribes a target range for communication efficiency, say $[0.8,0.9]$ (see Section \ref{sec:target}). At runtime, Alya continuously inquires the measured communication efficiency $\CE$ from TALP. If this one falls inside the target range, then the simulation proceeds without any change. If $\CE$ is lower than the minimum of the prescribed range, this means that the execution is not efficient enough and resources should be decreased. Employing a simple model, explained in Section \ref{sec:model}, Alya then estimates the number of cores needed to recover the target range. On the contrary, if the measured $\CE$ is above the range maximum, resources can be extended, similarly, they are removed. Once the new amount of resources is available, Alya writes restart files on disk, it is relaunched with a new partitioning, reads the restart files, and resumes the simulation. To ensure restart files can be written and read disregarding the number of subdomains, a specific I/O strategy is explained in Section \ref{sec:io}.\\

{\em Efficiency}. This paper presents an autonomous elastic computing methodology to achieve efficient CFD simulations, based on obtaining parallel efficiency metrics at runtime that allow us to estimate the optimum number of computing resources to use. 
This approach intends to trade-off power consumption, through controlling the resources to reach a target parallel efficiency, and performance. It thus belongs to the family of energy-to-solution techniques, as opposed to time-to-solution techniques. 
Traditionally, energy-to-solution techniques have been based on hardware parameter tuning, like CPU core frequency, CPU uncore frequency (e.g. cache, memory controller), or the number of OpenMP threads \cite{1559953,doi:10.1177/1094342018798452,10.1007/978-3-319-97136-0_11,1559985}. Application parameters tuning has also been investigated for example in \cite{chowdhury_anamika_2017_815852,venkatesh_kannan_2019_2808081} in the context of the READEX European project. Our strategy is a resource tuning strategy for MPI applications.\\

{\em Generic performance analysis}. Parallel efficiency, as well as software or hardware counters in general can be obtained with libraries.
Tracing tools like Extrae \cite{extrae-web}, ScoreP \cite{10.1007/978-3-642-31476-6_7} or HPCToolkit \cite{https://doi.org/10.1002/cpe.1553}, collect information such as PAPI counters, MPI, and OpenMP calls that happen during the execution and store them in a trace together with timing information. These traces can then be visualized and analyzed post-mortem with visualization tools available within the library or by specific tools like Paraver \cite{paraver} or Scalasca \cite{https://doi.org/10.1002/cpe.1556}. The trace-based tools provide detailed information on the execution after the application has finished and usually a performance analyst is required to process the data.
LIKWID \cite{liwkid} is a command-line performance tool suite, which, among other functionalities, provides hardware performance counters. The different metrics collected can be displayed at runtime with one of the utilities provided in the suite. However, these metrics do not include any parallel efficiency information, and the user needs to interpret them.
Also, the lightweight mpiP library delivers specific statistical information on MPI calls \cite{mpip} upon finalization of the execution. In this work, we rely on the library TALP, which provides parallel efficiency metrics at runtime. The metrics collected by TALP are defined in the PoP performance model \cite{wagner}, developed by the researchers of the European Centre of Excellence {\em Performance Optimisation and Productivity} \cite{coe-pop}; the PoP metrics are a set of efficiency and scalability indicators that can be obtained for MPI applications. In our case, the efficiency metrics reported by TALP allow the user to obtain the parallel efficiency, which is split into communication efficiency and load balance. This allows to compute the optimum number of the computational resources that must be used in the execution to achieve a given parallel efficiency defined by the user.\\


{\em Performance analysis of CFD codes}. As aforementioned, the calculation of the parallel efficiency of simulation codes usually relies on post-mortem analysis. In \cite{Gasulla20}, post-mortem analyses are performed on three different CFD codes using several performance tools. The PoP center of excellence \cite{coe-pop} offers continuous performance analysis of simulation codes; see for example \cite{opf} for OpenFoam or \cite{avbp2} for AVBP. It is common practice for High-Performance Computing (HPC) users to extrapolate parallel efficiency from computed speedup. To this aim, the speedup is obtained by timing executions of the code on a different number of cores and computing it based on timings for the lowest number of cores. Finally, this parallel efficiency is calculated using the ratio between the measured speedup and the ideal speedup (perfect speedup meaning efficiency of one) \cite{ueabs2,avbp}. Consequently, these measures are relative and not absolute. Moreover, they provide information on the timing but not on the parallel efficiency achieved. On the contrary, the TALP library considered in this work provides runtime measures of the different metrics composing the real and absolute parallel efficiency achieved in an execution.\\

{\em Elasticity}. Once the CFD code is adequately instrumented for runtime measures, dynamic resources allocation can be put in place to provide elasticity to the execution.
The elasticity term has become very popular with the evolution of Cloud Computing~\cite{cloud}. Cloud providers and software vendors have developed services to allow users to define rules to automatically scale up/down resources of their services when a certain metric reaches a threshold or a certain event is triggered~\cite{qu2018auto}. 
These auto-scaling services aim at automatically adjusting the required resources to the application load, and 
maximize the benefit
of the cloud pay-per-use model. For HPC systems, where the infrastructure is static but its workloads are dynamic, the elasticity concept has focused on supporting malleable jobs, i.e. applications that support changing the computing resources at runtime.
The research on this area has focused on three aspects: i) enabling malleability of jobs using resource managers such as SLURM \cite{yoo2003slurm,slurm} and OAR \cite{oar}; ii) scheduling of malleable jobs \cite{fotakis_et_al:LIPIcs:2019:11232,jansen2006approximation} and iii) the communication between the application and resource managers \cite{drom,malleability_mpi}. The present work thus proposes an original application of elastic computing to optimize a CFD execution in terms of parallel efficiency. \\

In the following four sections, we introduce the different components of the workflow, namely Alya, TALP, and COMPSs. We then describe the complete workflow in Section \ref{sec:compss}. Finally, the proposed strategy is validated by applying the proposed strategy to a series of CFD simulations in Section \ref{sec:results}.

\section{CFD code: Alya} 
\label{sec:alya}

\subsection{Physical and Numerical Modeling}

We consider in this paper the incompressible Navier-Stokes equations as a use case, but the strategy is extensible to any set of partial differential equations (PDE). We will assess two numerical schemes based on the finite element method, in which stabilizations depend on the state of the flow under consideration. For laminar to slightly turbulent flows, we rely on a Variational MultiScale Method (VMS) \cite{GHouzeaux_JPrincipe08} to stabilize both convection and pressure. The resulting system is then solved by extracting the pressure Schur complement and eventually converges to the monolithic solution \cite{GHouzeaux_RAubry_MVazquez09}. When considering fully turbulent flows, we rather opt for a fractional step scheme with a low dissipation scheme (EMAC scheme for convection \cite{CHARNYI2017289}) and Large Eddy Simulation (LES) turbulence modeling \cite{lehmkuhl18b}.
The main characteristics of both strategies are summarized in Table \ref{tab:fe}.
\begin{table}[htbp]
\centering
\begin{tabular}{lll} 
 \hline
                           & Implicit scheme                                        & Explicit scheme   \\
 \hline                                                
 Application range & Laminar-slightly turbulent & Fully turbulent \\
 Solution strategy & Pressure Schur complement \cite{GHouzeaux_RAubry_MVazquez09} & Fractional step \cite{codina} \\
 Time integration & Second order BDF & $4^{th}$ order Runge-Kutta \cite{10.1016/j.jcp.2016.10.040} \\
 Turbulent modeling & ILES & Vreman \cite{VRE04b-A} \\
 Wall treatment & No-slip & Law of the wall \cite{Owen18b} \\
 Convection stab. & VMS \cite{GHouzeaux_JPrincipe08} & None \cite{lehmkuhl18b} \\
 Pressure stab. & VMS \cite{GHouzeaux_JPrincipe08} & Continuous-discrete Laplacian \\
 Momentum solver & GMRES & Explicit \\
 Pressure solver & Deflated CG \cite{RLohner_FMut10}+linelet \cite{Soto} &   Deflated CG+linelet \\
 \hline
\end{tabular}
\caption{Two finite element schemes for the incompressible Navier-Stokes equations.} 
\label{tab:fe}
\end{table}
The objective of considering these two discretization schemes is to investigate the 
optimization of the parallel efficiency in two common scenarios, with different 
MPI communication patterns.
On the one hand, the monolithic approach requires solving two algebraic systems for the momentum and pressure equations, thus involving a large number of communications. On the other hand, the weight of the equation assembly compared to communications of the fractional step method is relatively higher than for the implicit scheme. The two schemes have therefore different patterns and 
require both to be treated by the proposed elastic approach.

\subsubsection{Parallelization and I/O}
\label{sec:io}

The parallelization of Alya is extensively described in \cite{Vazquez15d} for multi-core supercomputers, and in \cite{borrell20} for hybrid supercomputers including GPU accelerators. A dynamic load balance strategy based on OpenMP at the intra-node level is presented in \cite{lewi2}. In the present work, pure MPI parallelization is considered. One MPI process is referred to as the master process and is exclusively in charge of small outputs (basically convergence residuals and timings).

For the proposed strategy to be efficient, the overall simulation workflow must be parallel, as the CFD code is restarted when a different core count is required. If any sequential bottleneck remained in the restarting process of the code, it would cancel out the benefits of reallocating resources. 

To be able to write and read restart files independently of the number of subdomains, Alya relies on a unique entity numbering. In the finite element context, mesh entities are nodes, elements, and boundaries, on which values are susceptible to be required (e.g. primary variables). To this end, the initial mesh numbering provided by the mesher is used (see e.g. the strategy described in \cite{doi:10.1080/10618562.2020.1810676}). When a restart file needs to be written, an online data redistribution is performed based on the original numbering, then MPI I/O \cite{Prost2011,58c0da21e5b14ad1a57a79597b1ff9cb} is used to write contiguous chunks of data in parallel \cite{fournier2015recent,kodavasalperformance,jansen2017extreme}. When reading the restart files the inverse process is applied: first contiguous chunks of data are read in parallel, then a redistribution is used to construct fully defined subdomains for each parallel process.

\begin{figure}
  \centering
  \includegraphics[width=0.95\textwidth]{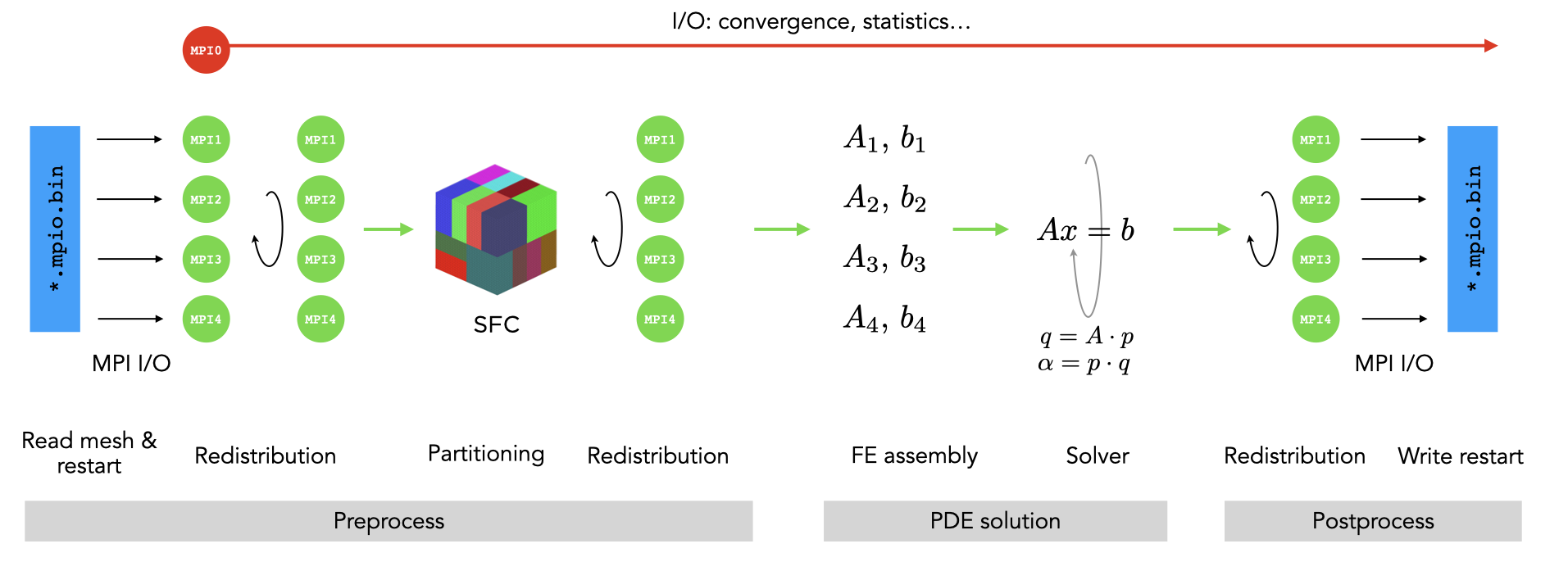}
  \caption{Alya parallel workflow.}
  \label{fig:alya}
\end{figure}

The complete parallel workflow of Alya is depicted in Figure \ref{fig:alya}. 
As aforementioned, the mesh is written following a global numbering that will be used throughout the simulation. Upon starting Alya, the mesh and restart files are read in parallel. Data is subsequently distributed to get local complete subdomains on each MPI process. Using this distributed data, parallel partitioning is carried out using a Space-Filling Curve (SFC) method \cite{Borrell18b}, which requires the coordinates of the element centers. Then, according to the element-to-MPI assignation given as an output of this partitioning, data is redistributed and the PDE solution strategy can start. Once Alya is required to write down the restart files, a new redistribution occurs and the data is written on disk using the initial global mesh numbering. 

\subsubsection{Two test cases}
\label{sec:validation}

\begin{figure}
  \centering
  \includegraphics[width=0.99\textwidth]{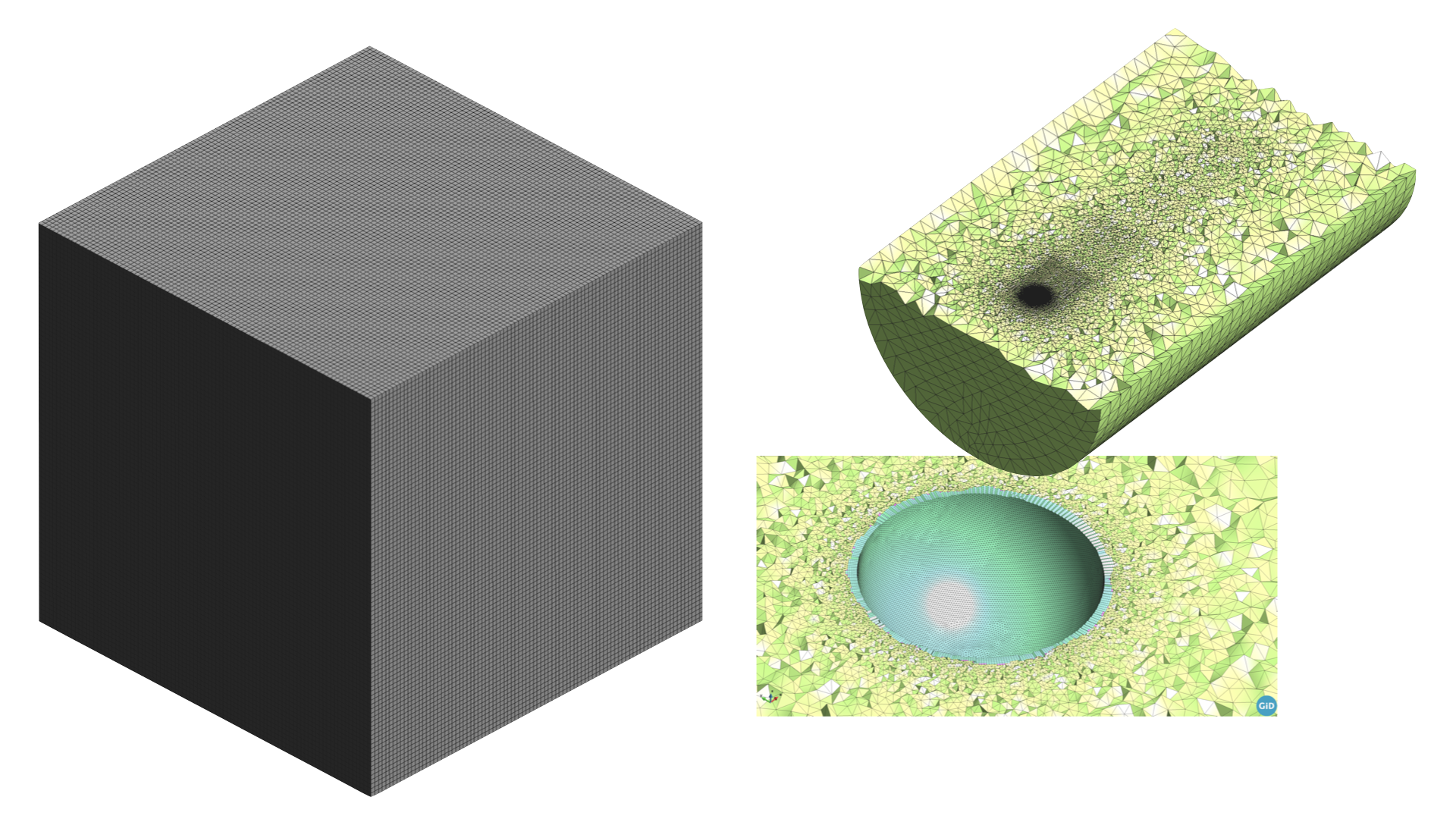}
  \caption{Mesh1 and Mesh2 used for the implicit and explicit schemes, respectively.}
  \label{fig:meshes}
\end{figure}

For the sake of validation, we will work with two different meshes and settings, for the implicit and explicit schemes respectively.
The characteristics of both meshes are listed in Table \ref{tab:meshes}. The first mesh is composed exclusively of hexahedra elements (HEX08), while the second one is hybrid, involving tetrahedra (TET04), prisms (PEN06), and pyramids (PYR05) that ensure a conformal transition from tetrahedra to prisms.
Global overview and details on the meshes are given in Figure \ref{fig:meshes}.
\begin{table}[htbp]
\centering
\begin{tabular}{lll} 
 \hline
                           & Mesh1                                        & Mesh2   \\
 \hline                                                
 Scheme tested & Implicit & Explicit \\
 Number of elements & 729k & 3.2M \\
 Elements & HEX08 & TET04, PYR05, PEN06 \\
 \hline
\end{tabular}
\caption{Meshes used for validation for the implicit and explicit schemes.} 
\label{tab:meshes}
\end{table}

On the one hand, on Mesh1, a cavity flow at Reynolds $10^2$ is solved using the implicit approach summarized in Table \ref{tab:fe}. On the other hand, on Mesh2, the turbulent flow over a sphere is solved at Reynolds $10^4$, using the Vreman turbulence model, as summarized in Table \ref{tab:fe}.

\section{Measuring Parallel Efficiency: TALP} 
\label{sec:talp}

The CFD code Alya relies on the library TALP to measure at runtime the parallel efficiency of the simulation. We now explain in detail how TALP works and the different metrics it measures and can provide to Alya.

\subsection{The library}

TALP \cite{talp} is a lightweight and scalable tool for online parallel performance measurement integrated within the Dynamic Load Balancing Library (DLB) \cite{LeWI,DLB} library. DLB is a framework that aims at improving the performance of parallel applications and is designed in a modular way. It is transparent and non-intrusive for the application and user, and it is integrated with MPI, OpenMP \cite{omp5} and OMPSs API's \cite{ompss}.

DLB includes three modules independent but coordinated and compatible: 

\begin{itemize}
    \item LeWI (Lend When Idle): provides a dynamic load balancing algorithm for hybrid applications.
    \item DROM (Dynamic Resource Ownership Management): offers a mechanism for dynamic resource management.
    \item TALP (Tracking Application Low-level Performance): measures parallel efficiency at runtime.
\end{itemize}

The TALP module is a profiling tool that collects statistical information about the contribution of every thread and process to the application execution. The tool measures load imbalance and communication efficiency by profiling each of the supported parallel programming models. These performance measures can be obtained at runtime through the use of an API, as well as from a post-mortem 
report. In the current work, we take advantage of the API offered to collect efficiency metrics at runtime. 
The measures done by TALP are based on defining two main states for a running process: useful work and communication. Based on these measures, it can provide parallel efficiency metrics as defined by the PoP centre of excellence \cite{pop,pdp,coe-pop}.

By adding the required calls inside Alya through the API, we are thus able to obtain the performance measurements during its execution (see Section \ref{sec:api}). Then, according to the efficiency values inquired by Alya, resources will be adjusted to fulfill an efficiency criterion. 

\subsection{Efficiency Metrics}

\begin{figure}[htbp!]
  \centering
  \includegraphics[width=0.7\textwidth]{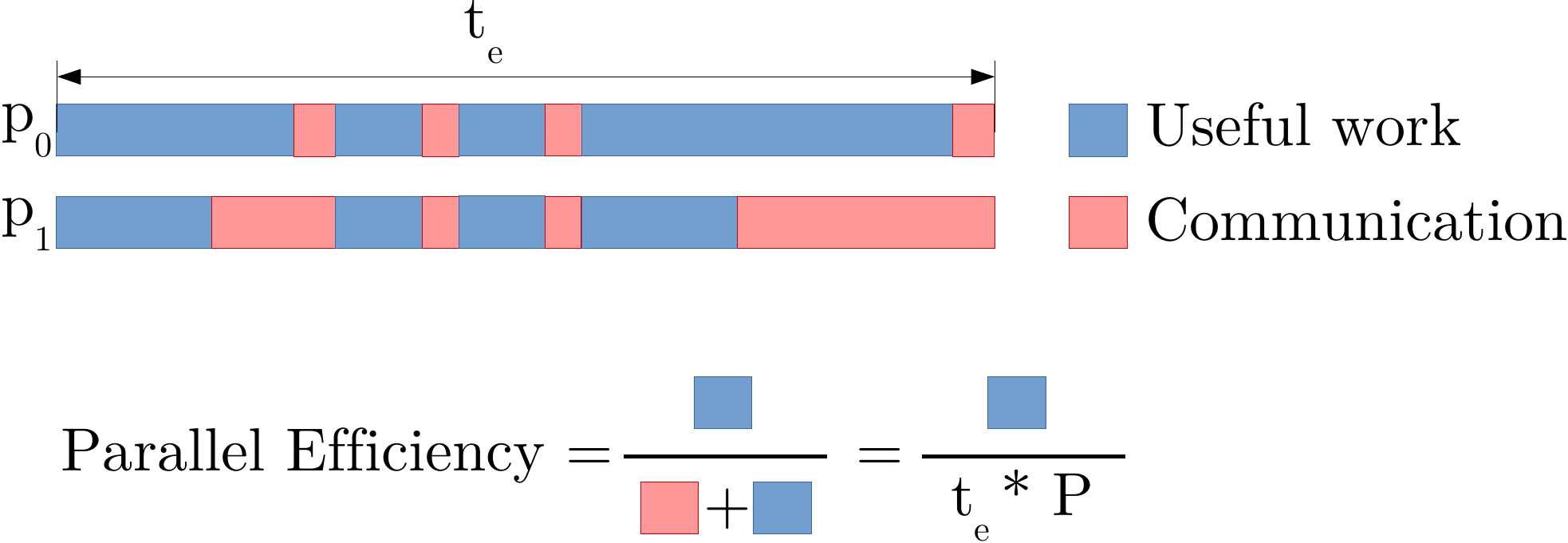}
  \caption{Graphical definition of Parallel Efficiency. $P$ is the number of available cores and $t_e$ the elapsed time.}
  \label{fig:PE}
\end{figure}

In this section, we explain the way the efficiency metrics are computed based on the profiling measurements.
As mentioned previously, we simplify the status of a process into two states, namely useful work and communication. Figure~\ref{fig:PE} shows a graphical representation of two processes running in parallel ($p_{0}$ and $p_{1}$) by coloring the two possible states.

In the HPC community, parallel efficiency is computed as the ratio between the useful computation and the total consumed resources.
Let us define $t_w^i$ and $t_c^i$ the useful work and communication times of process $i$ running on $P$ cores. The total elapsed time $t_e$ of an application is thus:
\begin{EQ}
  t_e = \max_i(t^i_w+t^i_c). \label{eq:elapsed_time}
\end{EQ}
We also introduce $t_w$ as the total useful work time of the application:
\begin{EQ}
  t_w = \sum_i t_w^i.
\end{EQ}
Accordingly, the Parallel Efficiency (PE) can be computed as:
\begin{EQ}
\PE = \frac{t_w}{t_e \times P} \label{eq:pe}
\end{EQ}

\begin{figure}[htbp!]
  \centering
  \includegraphics[width=0.7\textwidth]{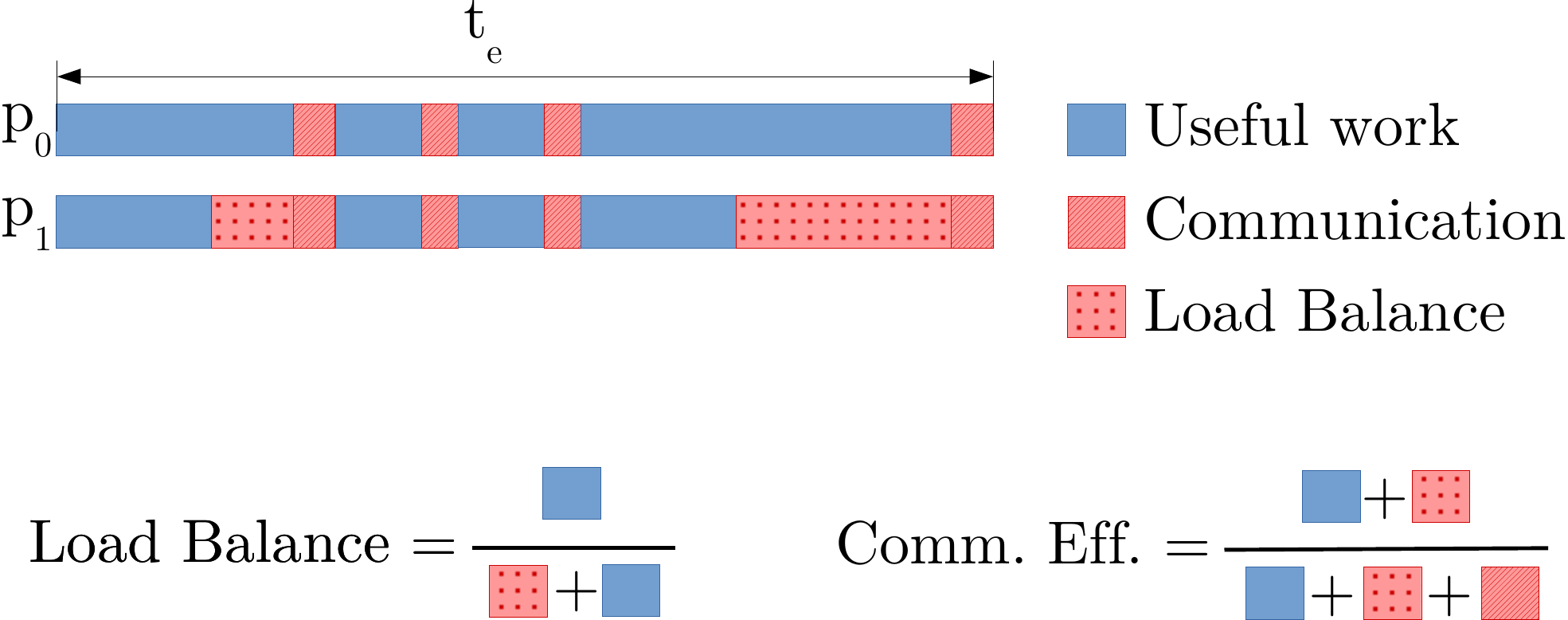}
  \caption{Graphical definition of Load Balance ($LB$) and Communication Efficiency ($CE$).}
  \label{fig:LB_CE}
\end{figure}

By using the PoP metrics, we divide the communication time into two separated metrics:
\begin{itemize}
    \item Load Balance (LB): time spent waiting for the most loaded process to complete its task.
    \item Communication Efficiency (CE): time spent in the MPI library, due to the inherent overheads of communication.
\end{itemize}
Figure~\ref{fig:LB_CE} shows a graphical representation of the difference between the two metrics.

Based on this definition, they can be computed as:
\begin{EQA}[rcl]
\CE &=& \frac{\max_i(t^i_w)}{t_e} \label{eq:ce}, \\
\LB &=& \frac{t_w}{\max_i(t^i_w) \times P}, \label{eq:lb} \\
\PE &=& \LB \times \CE.
\end{EQA}

This difference is important because the approach to address a load balance issue is different from the one to address a communication efficiency issue. On the one hand, a low $LB$ can be solved with a better domain partition or a dynamic load balancing mechanism. On the other hand, a low $CE$ is usually an indicator that too many resources are used for the current size of the problem. All the previous definitions are summarized in Table \ref{tab:perf}.
\begin{table}[htbp]
\centering
\begin{tabular}{lll} 
 \hline
  Definition & Symbol       & Value    \\
 \hline                                                
 Work time of $i$          & $t_w^i$ & Measured \\
 Communication time of $i$ & $t_c^i$ & Measured \\
 Total work                & $t_w$   & $t_w = \sum_i t_w^i$ \\
 Elapsed time              & $t_e$   & $t_e = \max_i(t_w^i + t_c^i)$ \\
 Load balance              & $\LB$   & $\LB = t_w / (\max_i(t_w^i) \times P$) \\
 Communication efficiency  & $\CE$   & $\CE = \max_i(t_w^i)/t_e$ \\
 Parallel efficiency       & $\PE$   & $\PE = \ t_w / (t_e \times P)$ \\
 \hline
\end{tabular}
\caption{Definitions and performance metrics. $i$ refers to MPI process $i$.} 
\label{tab:perf}
\end{table}

\subsection{Implementation of the API in Alya}
\label{sec:api}

TALP provides a Fortran 2008 interface. To activate the 
measurements the following sequence is necessary. First, two types should be declared, as 
outlined
in Listing \ref{lst:declaration}.
The next step consists in registering the monitoring,
as outlined
in Listing \ref{lst:register}.
Then, to open and close a region to be monitored, the code shown in Listing \ref{lst:measure} should be implemented.
Finally, Listing \ref{lst:timings} shows how to get the timings in nanoseconds.
\begin{lstlisting}[caption={Declaration of TALP counters in Alya.},label={lst:declaration},language={[95]Fortran}]
use, intrinsic :: ISO_C_BINDING
include 'dlbf_talp.h'
type(dlb_monitor_t), pointer :: dlb_monitor
type(c_ptr)                  :: dlb_handle
\end{lstlisting}
\begin{lstlisting}[caption={Registering of TALP counters in Alya.},label={lst:register},language={[95]Fortran}]
dlb_handle = DLB_MonitoringRegionRegister(c_char_"my region"//C_NULL_CHAR)
\end{lstlisting}
\begin{lstlisting}[caption={Performance measurements with TALP counters in Alya.},label={lst:measure},language={[95]Fortran}]
integer(4) :: ierr
ierr = DLB_MonitoringRegionStart(dlb_handle)
(... region to be monitored ...)
ierr = DLB_MonitoringRegionStop(dlb_handle)
\end{lstlisting} 
\begin{lstlisting}[caption={Obtaining the TALP counters in Alya.},label={lst:timings},language={[95]Fortran}]
real(8) :: elapsed_time
real(8) :: accumulated_MPI_time
real(8) :: accumulated_comp_time

call c_f_pointer(dlb_handle, dlb_monitor)
elapsed_time          = real(dlb_monitor % elapsed_time,8)
accumulated_MPI_time  = real(dlb_monitor % accumulated_MPI_time,8)
accumulated_comp_time = real(dlb_monitor % accumulated_computation_time,8)
\end{lstlisting} 

\section{Workflow manager: COMPSs} 

We have just presented the CFD code and the library TALP to measure the performance at runtime. We now briefly introduce the workflow manager, COMPSs.
COMPSs \cite{compss_softwareX} is a task-based programming model and runtime for simplifying the development and execution of parallel workflows in distributed computing environments. From an annotated sequential code, the COMPSs runtime can detect data dependencies between the defined tasks' invocations, creating a task-dependency graph that allows the inference of the application's inherent parallelism. Based on the required parallelism in each computation phase, the COMPSs runtime interacts with the resource manager, such as SLURM, to adapt the computational resources to the application needs \cite{compss_servicess}. The dynamic job capability of SLURM, which allow us to increase or reduce the resources assigned to a job in execution, is described in 
\cite{slurm-dynamic}.
This feature ensures that new resources in the extended job are accessible by the Process Management Interface (PMI) of MPI. This strategy can be easily implemented in other schedulers that provide a similar feature. If the scheduler does not provide this feature, COMPSs can also perform the scaling by adding new resources to the application through standard job submission. However, this option will only work if the scheduler allows starting an MPI application in resources from different jobs.

In the present case, COMPSs is used to implement and execute the optimization workflow that orchestrates the Alya execution. The Alya executions are defined as COMPSs tasks. A COMPSs workflow is defined to spawn the Alya computations collecting its resource requirements through interaction with the resource manager and restarting the Alya execution with the new resource configuration and the minimum downtime.

\section{Optimization workflow: elastic computing} 
\label{sec:compss}

The overall workflow combines a CFD code Alya, a performance library TALP to measure the communication efficiency $\CE$, and a workflow manager COMPSs. The optimization strategy consists in adjusting the resources according to a measured performance criterion, the communication efficiency $\CE$ herein. In practice, the communication efficiency is inquired by Alya to TALP at a given frequency, in terms of the number of time steps (e.g. each 10 time steps), to have a representative average. Once measured, if $\CE$ falls outside a prescribed range, $[\CE_{min},\CE_{max}]$ (e.g. $[0.8,0.9]$), then a new number of cores is estimated and requested. The estimation of the number of cores will be described in Section \ref{sec:target}.

\subsection{Software strategy}

Listing~\ref{lst:pycompss_workflow} shows the COMPSs workflow implemented to manage the execution of
the Alya computations and adapt the resources to its computational needs.
\begin{lstlisting}[caption={Code snippet of the adaptation workflow.},label={lst:pycompss_workflow},language={Python}]
#Alya task definition
@mpi(binary="/apps/Alya/bin/Alya.x",
     flags="-x LD_PRELOAD=/apps/dlb/lib/libdlb_mpif.so",
     processes=current_processes)
@task(returns=1, dataset=IN, comm=STREAM_OUT)
def alya_task(dataset=None, comm=None):
    pass

...
task_exit_value = STOPPED
while task_exit_value == STOPPED:
    # Create the stream to communicate resources estimations
    comm = DistroStream()
    # get current available 
    current_processes = compss_get_number_of_resources()
    #executes alya in a group called Alya
    with TaskGroup("Alya"):
        task_exit_value = alya_task(dataset, comm)
    while not comm.is_closed():
        estim_resources = comm.poll()
        new_resources = estim_resources - current_processes
        if new_resources > 0 :
            #Request resources and cancels Alya group tasks once ready
            compss_request_resources(new_resources, "Alya")
        elif new_resources < 0:
            #Cancel Alya group tasks and release resources
            compss_free_resources(new_resources*(-1), "Alya")
    #waits for Alya task end and syncronize the return code
    task_exit_value = compss_wait_on(task_exit_value)
\end{lstlisting}

\begin{figure}
  \centering
  \includegraphics[width=0.8\textwidth]{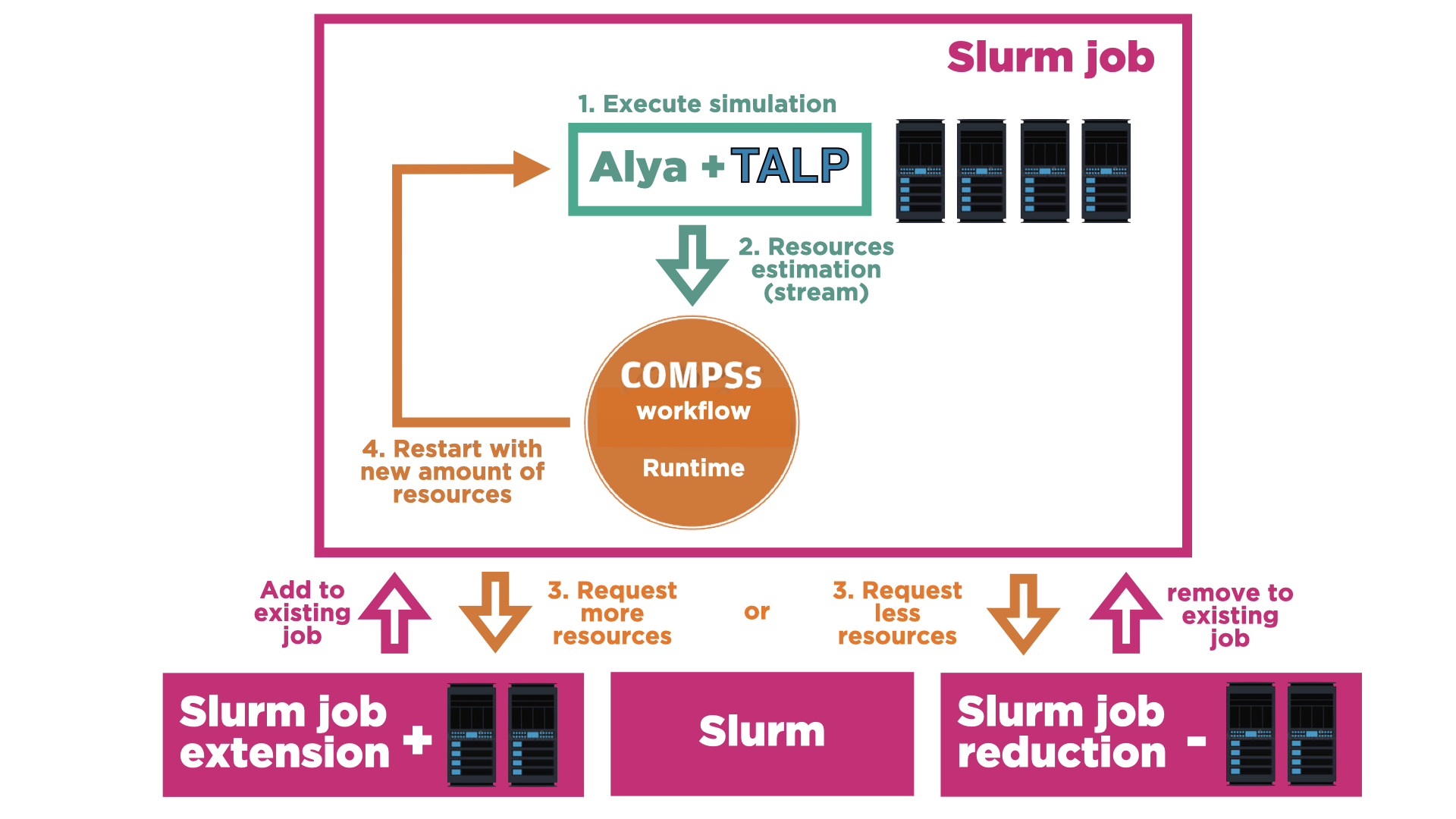}
  \caption{Optimization System Overview.}
  \label{fig:adaptation-system}
\end{figure}

The workflow defines a COMPSs MPI task, which starts the Alya computation using the available resources, and a stream to communicate the computational needs during the Alya execution based on target efficiency. The workflow consists of a while loop that starts the Alya computation and polls the stream to get a resource estimation according to the desired target efficiency. This estimation is compared to the current resources to decide if a resource increase or reduction is needed. To apply these actions, the COMPSs runtime interacts with the SLURM resource manager to extend or reduce the current job adding more resources or removing nodes to this job (released to other users), respectively. If the decision is to reduce the number of nodes, the COMPSs runtime selects a node to reduce, applies the reduction to SLURM, and restarts the Alya execution with the reduced resources. If the decision is to increase resources, COMPSs requests a job extension to SLURM and, once the new resource is ready, the Alya computation is restarted with the new resources. Otherwise, the computation continues until a new estimation is received or the computation finishes. The process is illustrated in Figure \ref{fig:adaptation-system}.

\subsubsection{Communication between Alya and COMPSs}

A two-way communication strategy is required between Alya and COMPSs. On the one hand, Alya should communicate to COMPSs when a new number of resources is required. This is achieved by writing this information at each time step into a file, read by COMPSs periodically. On the other hand, once the new resources are available, COMPSs should tell Alya to write the restart files and stop the run. This is achieved through a signal {\tt SIGTERM}:
\begin{EQ}
  \mbox{Alya} \stackrel[\mbox{File}]{\mbox{\tt SIGTERM}}{\lrarrow} \mbox{COMPSs}
\end{EQ}
Note that in the current implementation, until resources are not available, the simulation proceeds without checking a new communication efficiency value. Therefore the simulation is continuously running.

\subsubsection{Target range of communication efficiency}
\label{sec:target}

The global procedure to adjust the resources consists in estimating the number of cores $n^*$ to be used in the next run given the performance measure of the current run. While running on $n$ cores, Alya continuously obtains work and communication times from TALP. From these measures, the communication efficiency $\CE$ can be then computed using Equation \eqref{eq:ce}.
As mentioned previously, the user should prescribe a target range for the communication efficiency, within which the current resources are accepted. Let $\CE^*$ be the target efficiency defined as
\begin{EQ}
 \CE^* = \frac{1}{2} (\CE_{min}+\CE_{max})
\end{EQ}
one wishes to obtain running with $n^*$ cores (unknown), with work and communication times $t_w^{i*}$ and $t_c^{i*}$, respectively, for each core $i=1 \dots P$. A summary of the current and target run variables is given in Table \ref{tab:estimate}.
\begin{table}[htbp] \centering 
\begin{tabular}{lllcll} 
\hline
                        & \multicolumn{2}{c}{Current run} && \multicolumn{2}{c}{Target run}   \\
\cline{2-3} \cline{5-6}             
\mbox{Work time}        & $t^i_w$  & Measured             && $t^{i*}_w$ & Estimated  \\
\mbox{Comm. time}       & $t^i_c$  & Measured             && $t^{i*}_c$ & Estimated  \\
\mbox{Comm. efficiency} & $\CE$   & Computed               && $\CE^*$ & Prescribed  \\
\mbox{Number of cores}. & $n$     & Given                  && $n^*$   & Estimated  \\
 \hline
\end{tabular}
\caption{Using runtime measures $\CE$ to estimate the number of cores $n^*$ to obtain the user prescribed target efficiency $\CE^*$.} 
\label{tab:estimate} \end{table}

\subsubsection{A model for target communication efficiency}
\label{sec:model}

To close the workflow, we need to estimate the number of cores $n^*$ required to obtain $\CE^*$. To this end, we will make three hypothesis that we will verify on the Mesh1 simulation defined in Section \ref{sec:validation}.
\begin{figure}
  \centering
  \includegraphics[width=0.32\textwidth]{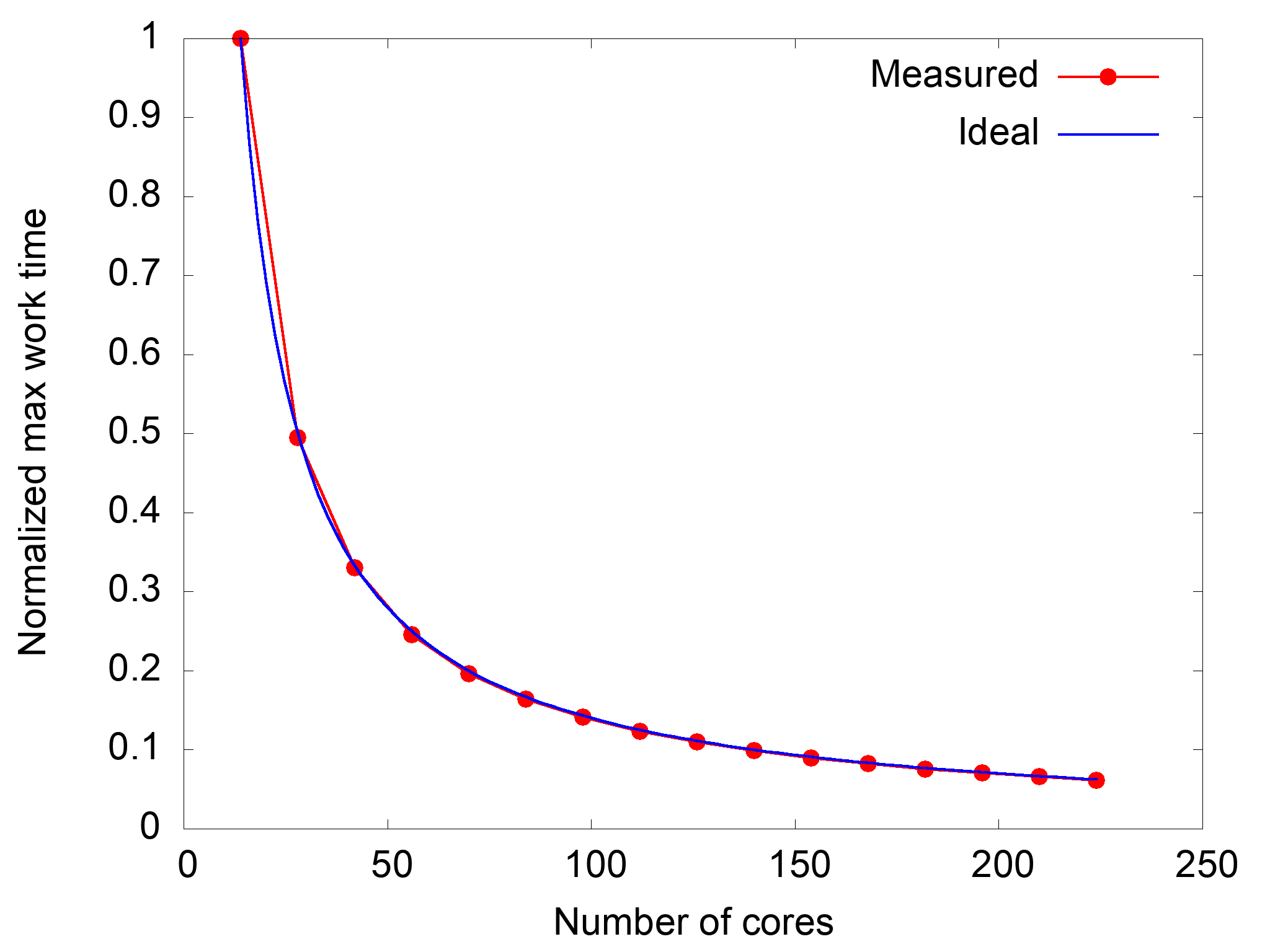}
  \includegraphics[width=0.32\textwidth]{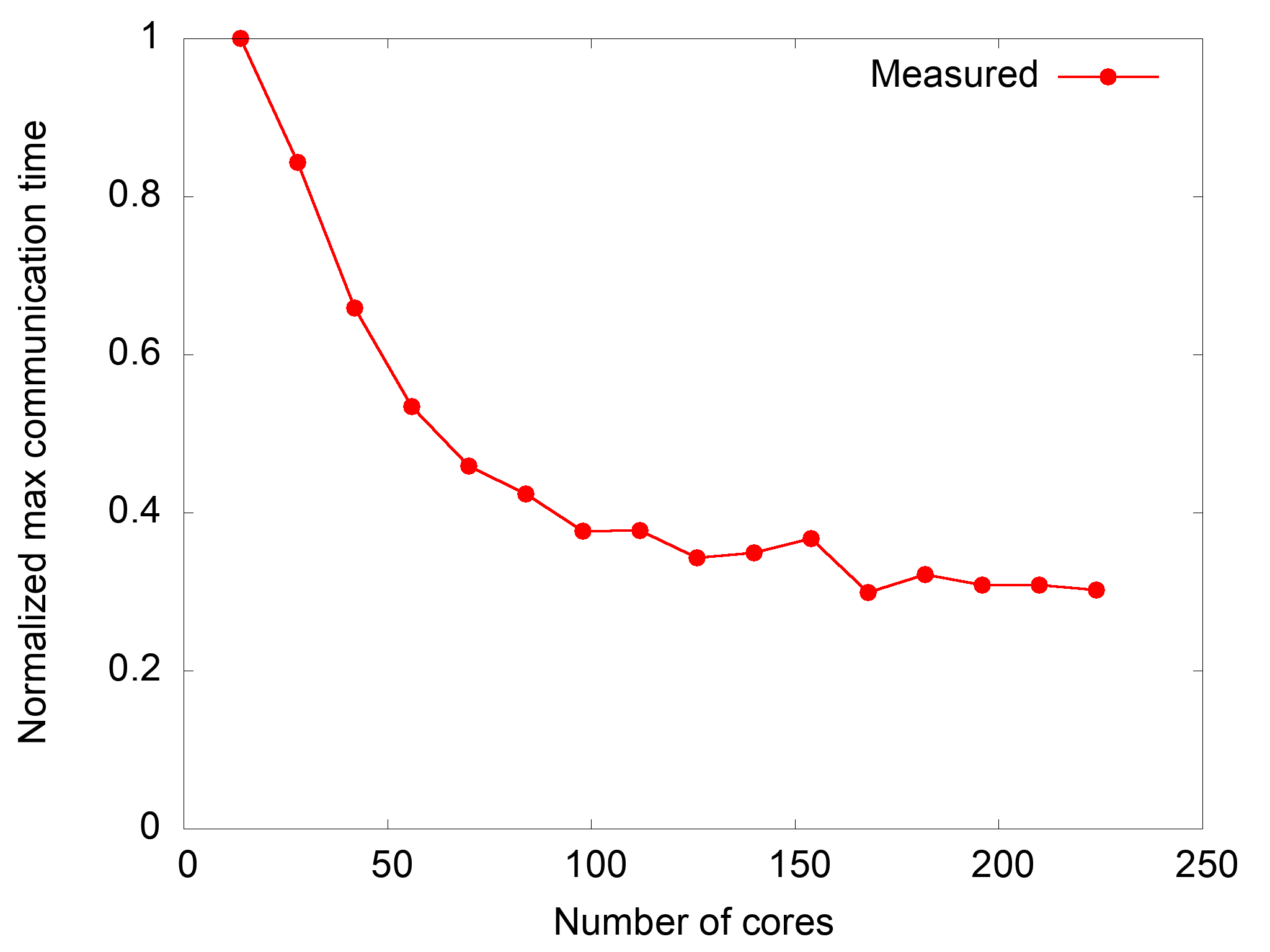}
  \includegraphics[width=0.32\textwidth]{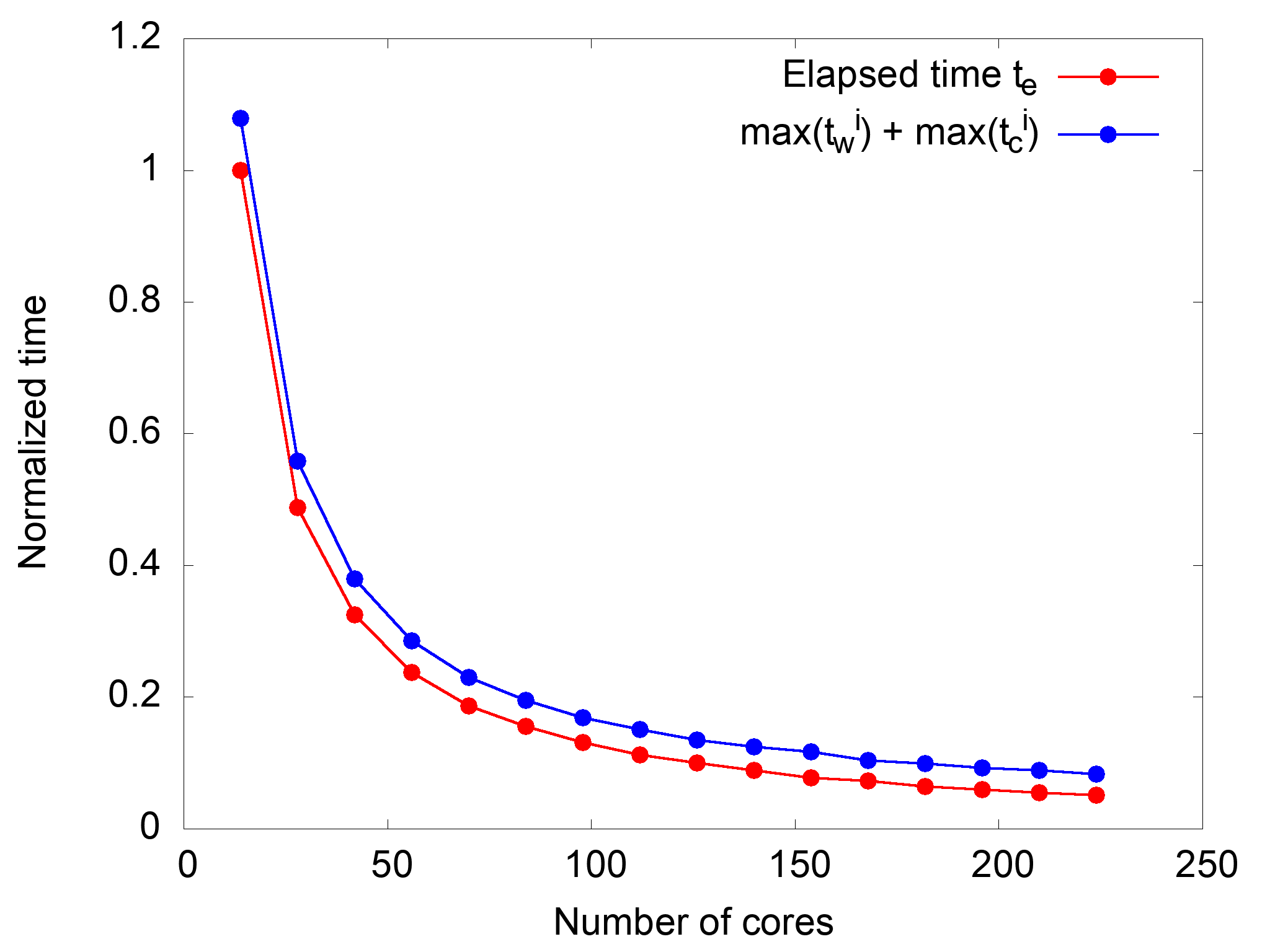}
  \caption{Checking the hypothesis for the estimate.
  (Left)  Work is perfectly scalable.
  (Mid.)  Maximum communication time is constant.
  (Right) Elapsed time can be approximated.
  }
  \label{fig:hyp}
\end{figure}

First, we assume that the work is perfectly scalable (perfect speedup) in average (Figure \ref{fig:hyp} (Left)), that is  
\begin{EQ}
  \max_i(t_w^{i*}) = \max_i(t^i_w) \, \frac{n}{n^*}. \label{eq:hyp1}
\end{EQ}
Then, we will assume that the maximum communication time is similar when changing the resources from $n$ to $n^*$, that is:
\begin{EQ}
  \max_i(t^{i*}_c) \approx \max_i(t^i_c). \label{eq:hyp2}
\end{EQ}
We see from Figure \ref{fig:hyp} (Mid.) that for a high number of cores this is almost true, although the approximation fails
for a small number of cores.
Finally, we will assume that the elapsed time defined in Equation \eqref{eq:elapsed_time} can be approximated by:
\begin{EQ}
 t_e \approx \max_i(t^i_w) + \max_i(t^i_c), \label{eq:hyp3}
\end{EQ}
although the right-hand side is strictly an upper bound. Figure \ref{fig:hyp} (Right) shows little difference between both.

Using Equations \eqref{eq:hyp1}, \eqref{eq:hyp2} and \eqref{eq:hyp3}, together with the definition of the communication efficiency given by Equation \eqref{eq:ce}, we obtain the number of cores $n^*$ required to attain efficiency $\CE^*$ as
\begin{EQ}
\mbox{Target number of cores:} \quad
 n^* = n \left( 1-\frac{1}{\CE^*}\right) \left( 1-\frac{1}{\CE} \right)^{-1}. \label{eq:nstar}
\end{EQ}

\begin{figure}
  \centering
  \includegraphics[width=0.75\textwidth]{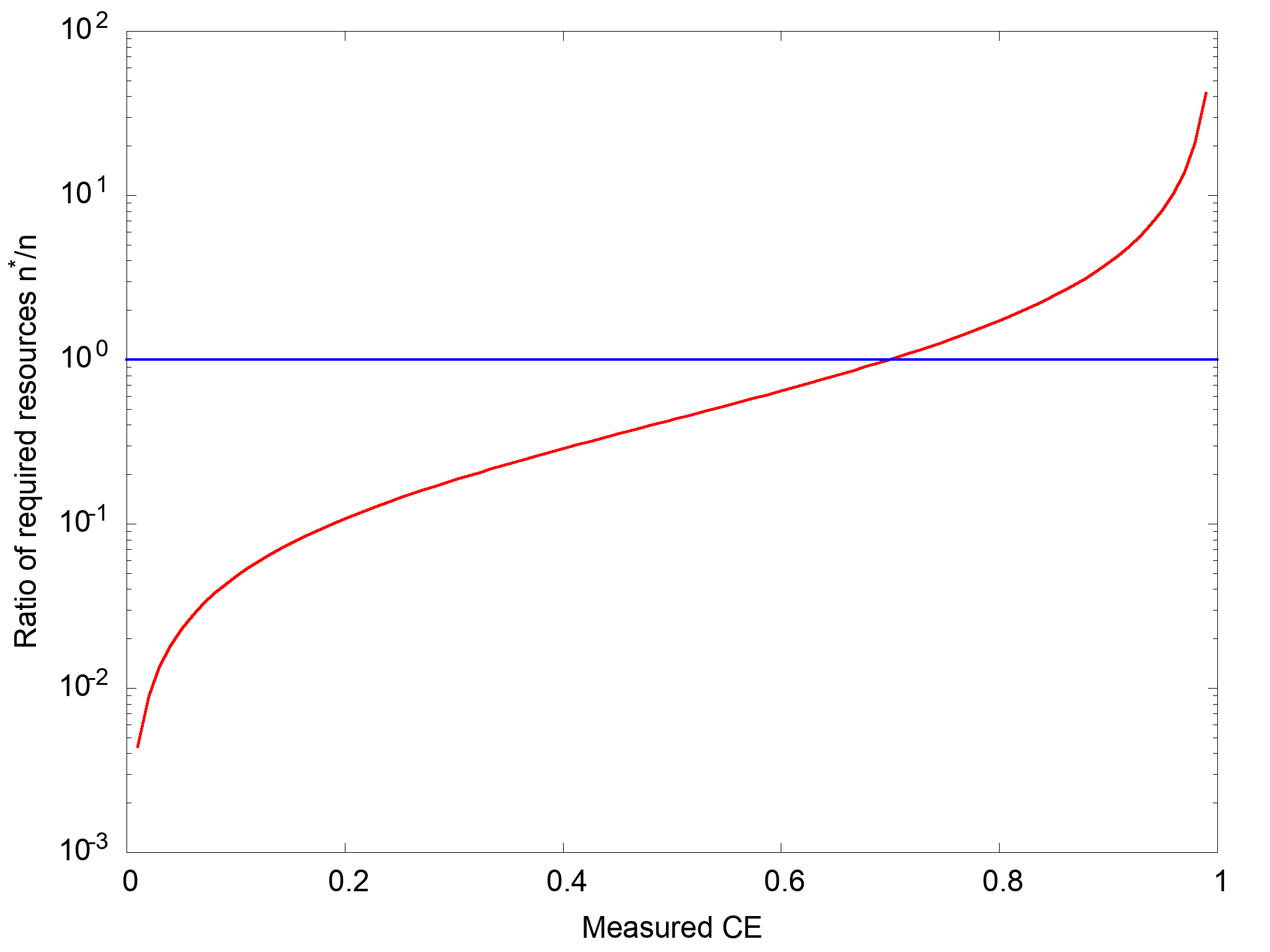}
  \caption{Resources to require depending on measured $\CE$, for a target $\CE^*=0.7$.}
  \label{fig:target}
\end{figure}

Figure \ref{fig:target} illustrates the formula for a target efficiency of $0.7$. If the measured communication efficiency is greater than 0.7, we fall above the horizontal line and the number of cores can be increased $(n^*/n>1)$. Conversely, if we measure a parallel efficiency lower than 0.7, the number of cores should be decreased to reach a higher efficiency $(n^*/n<1)$. 

\begin{figure}
  \centering
  \includegraphics[width=0.95\textwidth]{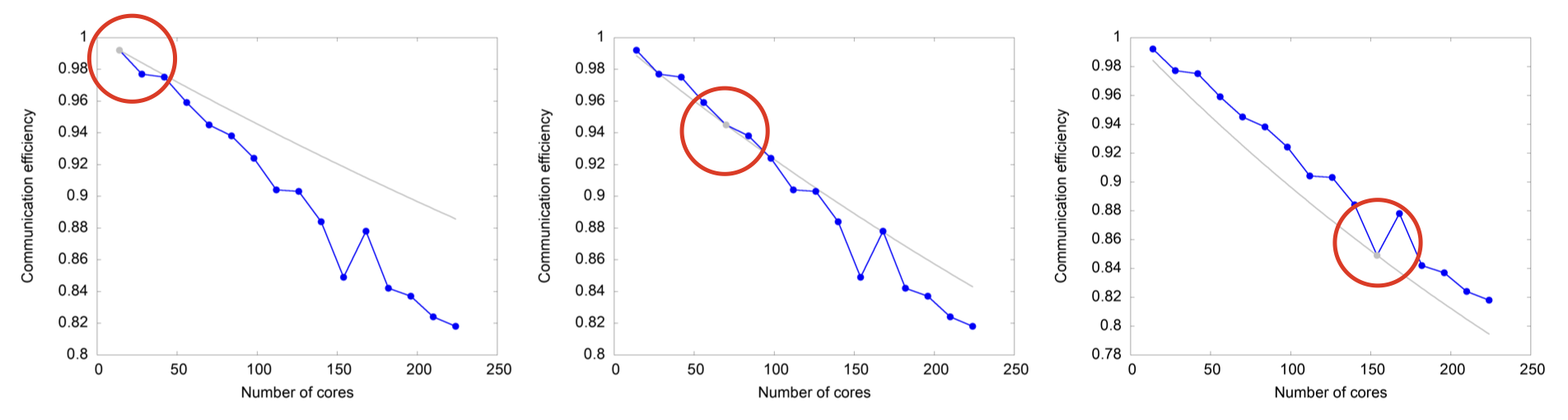}
  \caption{Prediction of Equation \eqref{eq:nstar2} to guess what would be the target $\CE^*$ for a given number of target cores $n^*$ compared to real measures. Three references for $n$ have been considered for the estimate.}
  \label{fig:predi}
\end{figure}

It is assumed that the estimate \eqref{eq:nstar} is used without having prior knowledge of the simulation communication properties. However, we will test it in the context of the Mesh1 example, to see if the approximation makes sense. For this we will revert the equation to find 
$\CE^*$ as a function of $n^*$:
\begin{EQ}
  \CE^* = \left[  1-\frac{n^*}{n}\left(1-\frac{1}{\CE}\right) \right]^{-1}. \label{eq:nstar2}
\end{EQ}
Figure \ref{fig:predi} shows the prediction of this equation for the target $\CE^*$ one would obtain as a function of $n^*$ compared to the real communication efficiency dependence on the number of cores $n^*$. 
We have considered three different reference numbers of cores $n$, each one corresponding to one of the plots.
We observe that locally, that is around a given $n/\CE$ pair identified by the red circle, the prediction is relatively good, except in the third case, for which a non-smooth dependence is measured. This suggests that the estimate should be applied carefully without requiring too big changes to the number of cores.

\subsubsection{Control parameters}

In this section, we will give all the control parameters required to set up an elastic simulation.
We already mentioned in Section \ref{sec:target} that the user should provide a target range. To obtain stable results on the communication efficiency (to avoid reacting to one-time events), we stated in Section \ref{sec:compss} that an averaging period would enable us to stabilize the predictions. Also, we have just seen in the previous section that it may be desirable to control the rate of change $r$ of the number of cores to make Equation \eqref{eq:nstar} valid. This rate of change limits the number of core $n_e^*$ estimated by Equation \eqref{eq:nstar} to:
\begin{EQ}
 \frac{n_e^*}{r} < n^* < n_e^* \, r
\end{EQ}

In addition, to control the effects of possible odd events, we will limit the range of the possible number of cores, based on experience, on the system, and the available number of cores. Also, a starting number of cores should be prescribed to start the first optimization step. Finally, a starting time step from which the measurements are activated will help to avoid a warm-up behavior of the execution. The following list summarizes the control parameters proposed in this work.   
\begin{itemize}
\item Communication efficiency target range;
\item Averaging period;
\item Rate of change of number of cores;
\item Minimum number of cores;
\item Maximum number of cores;
\item Initial number of cores;
\item Starting time step.
\end{itemize}                                       

\subsubsection{Overheads}

Let us comment on the possible overheads compared to a classical CFD simulation.
In the case of Alya, the overhead consists of restarting the CFD simulation. However, this restarting process is limited in terms of CPU time by the fully parallelized Alya workflow described in Section \ref{sec:io}. In addition, restarting is a common unavoidable operation carried out in CFD simulations to provide checkpointing, which costs is similar to the cost of a post-process output. Regarding the cost of a simulation, this extra cost can be neglected. 

Apart from the CFD code itself, the workflow involves two additional ingredients, namely TALP and COMPSs. On the one hand, TALP is integrated into Alya and tested regularly by the Alya performance suite \cite{perfsuite}. Overhead of a maximum of $3\%$ is obtained, which is reasonable given the potential gain expected by the workflow. On the other hand, the overhead of invoking a remote task with PyCOMPSs is about $10ms$, which is negligible compared with overall cost of a CFD simulation. An important overhead could be added by the management of the elasticity. There are two cases to consider, when requesting new resources and when releasing them. 
When requesting new resources, COMPSs submits new jobs to queueing system to access the missing resources. The submission of the job will take some milliseconds which can be also insignificant compared to the simulation time. However, the resource manager can take some time to provide the requested resources. To overcome these overheads, the workflow keeps the CFD simulation running with the old configuration until the new resources are available. In the case of releasing resources, COMPSs restarts the CFD with a smaller resource configuration and releases the other resources using the SLURM API which is performed asynchronously without adding extra overhead.

\section{Results}
\label{sec:results}

We now present the optimization results for both the implicit and explicit schemes. 
All the runs have been executed on Minotauro supercomputer, located at Barcelona Supercomputing Center. Minotauro's current configuration is:
\begin{itemize}
\item 38 bullx R421-E4 servers, each server with:
\begin{itemize}
\item 2 Intel Xeon E5–2630 v3 (Haswell) 8-core processors, (each core at 2.4 GHz,and with 20 MB L3 cache)
\item Peak Performance: 250.94 TFlops
\item 120 GB SSD (Solid State Disk) as local storage
\item 1 PCIe 3.0 x8 8GT/s, Mellanox ConnectX®–3FDR 56 Gbit
\item 4 Gigabit Ethernet ports.
\end{itemize}
\item The operating system is RedHat Linux 6.7.
\end{itemize}

For the whole optimization campaign, we give the control parameters in tables. In the figures, the shadowed rectangle indicates the target range; the blue line the number of cores, and the red line the measured communication efficiency. For all the simulations, the minimum number of cores is set to 15, equivalent to one node of the system, while the maximum number of cores is set to 240, equivalent to 16 nodes (we have found higher efficiencies using 15 cores per node instead of 16).

\subsection{Implicit scheme}

\subsubsection{Tests 1/2: stability of the optimization process}

We will start with two tests to analyze the evolution of the number of cores in the implicit case. Tables \ref{tab:exp1} and \ref{tab:exp2} show the set of control parameters.
\begin{table}[htbp] \centering \begin{tabular}{ll}  \hline
 Parameter                        & Value   \\
 \hline                                                
 Communication efficiency target range & $[0.9,0.92]$ \\
 Averaging period & 10 time steps\\
 Rate of change of number of cores  & 2 \\
 Initial number of cores & 15 \\
 Starting time step & 5 \\
 \hline
\end{tabular} \caption{Parameters for Test 1: starting with small number cores.} 
\label{tab:exp1} \end{table}
\begin{table}[htbp] \centering \begin{tabular}{ll}  \hline
 Parameter                        & Value   \\
 \hline                                                
 Communication efficiency target range & $[0.97,0.98]$ \\
 Averaging period & 10 time steps\\
 Rate of change of number of cores  & 2 \\
 Initial number of cores & 90 \\
 Starting time step & 5 \\
 \hline
\end{tabular} \caption{Parameters for Test 2: starting with large number of cores.} 
\label{tab:exp2} \end{table}
We are interested here in studying the stability of the optimization process by changing the initial number of cores. In Test 1, where we start with a few cores (15), the initial value of $\CE$ is high (around 0.98), so the number of cores will be increasing up to 116 cores until the target range is reached. We observe that this range is reached after three optimizations steps. On the other hand, we start Test 2 with a large number of cores so that the initial $\CE$ is below the target one (around 0.94). We observe that the target range is reached in two steps with 20 cores. In both cases, the algorithm is quite stable and converges quickly to the target range.
\begin{figure}
  \centering
  \includegraphics[width=0.49\textwidth]{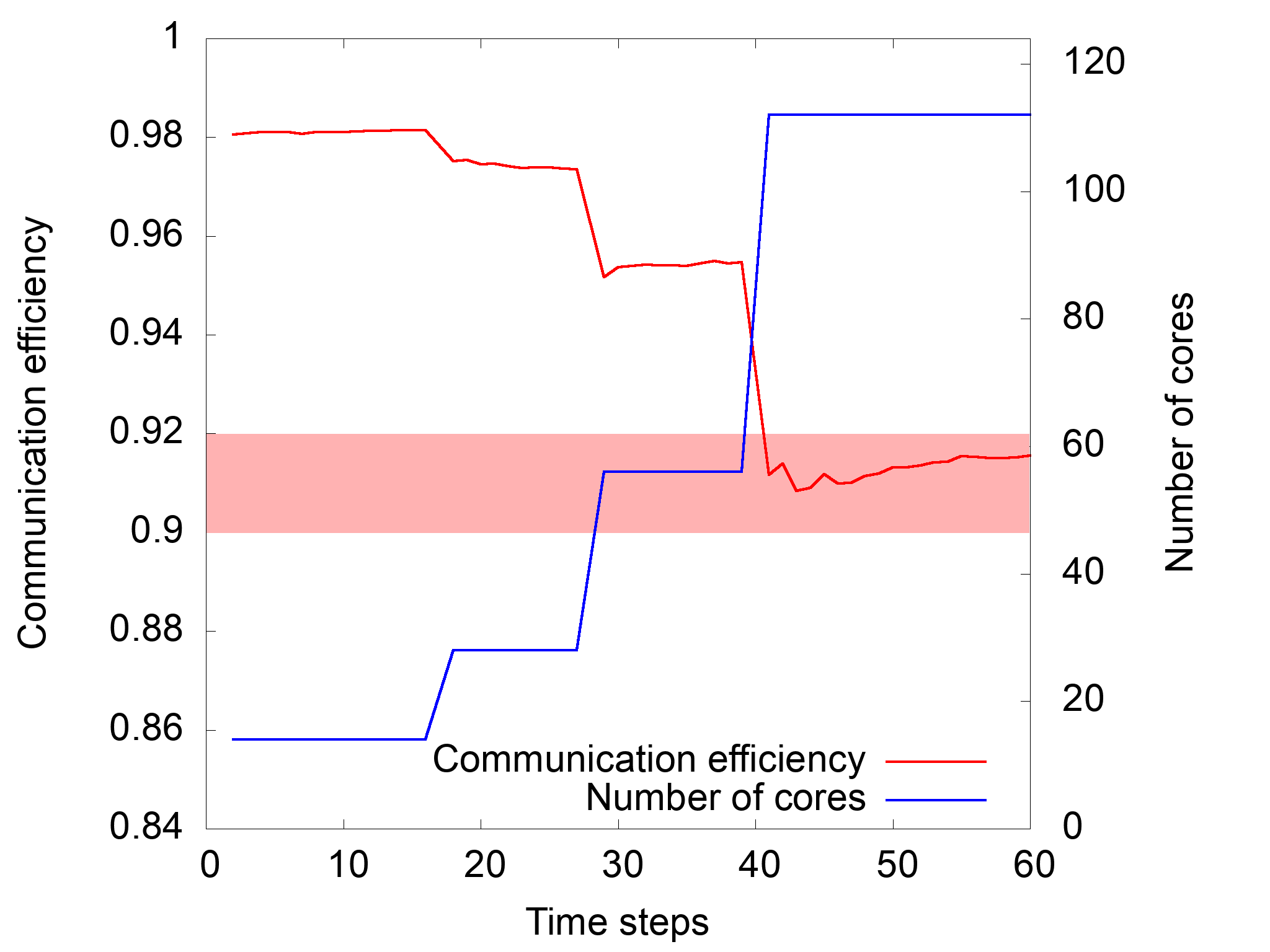}
  \includegraphics[width=0.49\textwidth]{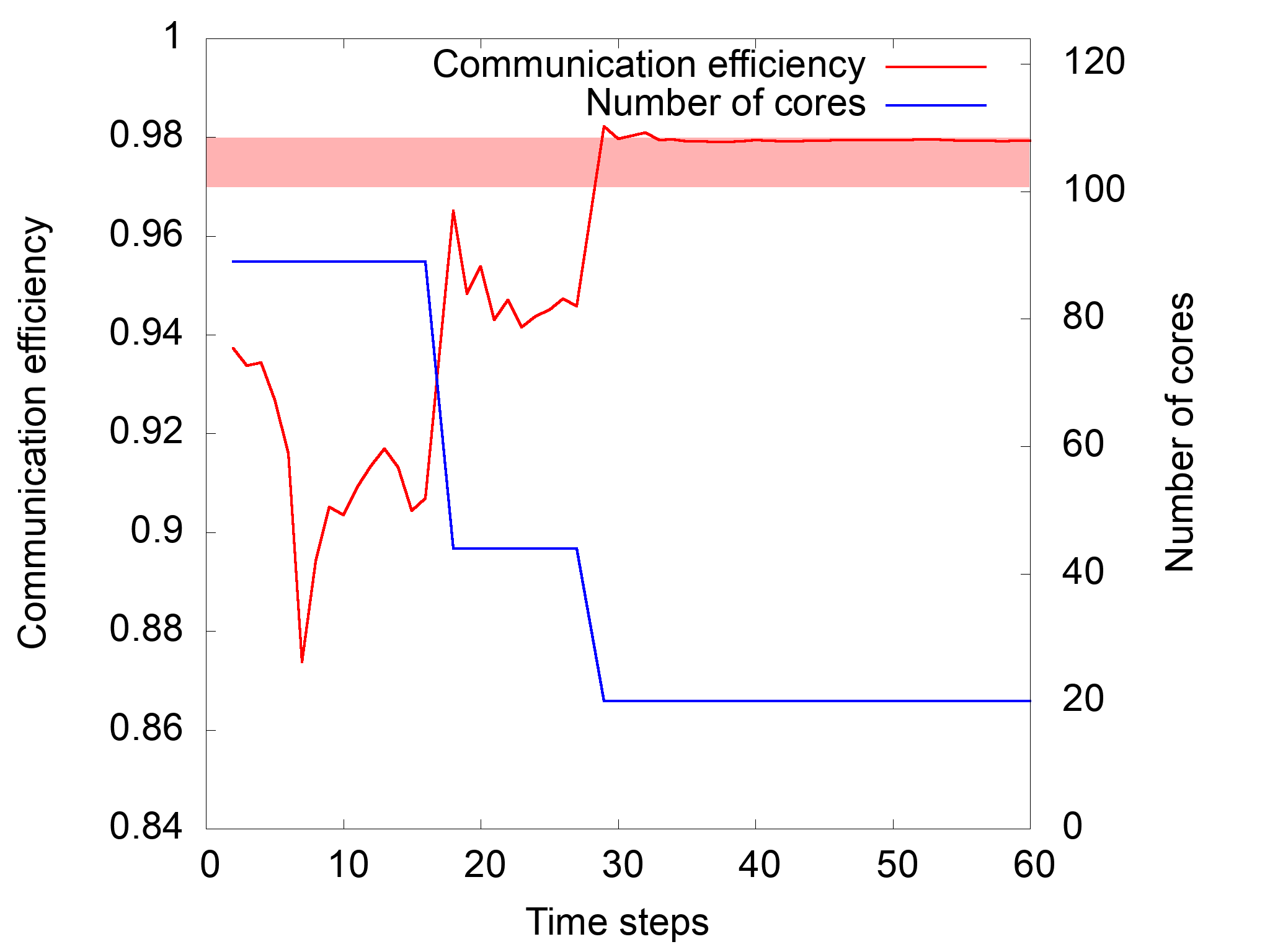}
  \caption{Optimization evolution in terms of $\CE$ and number of cores $n$ along the time integration.
  (Left)  Test 1: starting with a small number of cores. See Table \ref{tab:exp1}.
  (Right) Test 2: starting with a large number of cores. See Table \ref{tab:exp2}.
  }
  \label{fig:exp12}
\end{figure}

\subsubsection{Test 3/4: Rate of change}

We now study the effect of the rate of change. To start with a lower $\CE$ in the simulation, we increase the number of iterations of the iterative solvers and thus the relative weight of the MPI communications. 
Tables \ref{tab:exp3} and \ref{tab:exp4} show the set of control parameters used for these tests.
\begin{table}[htbp] \centering \begin{tabular}{ll}  \hline
 Parameter                        & Value   \\
 \hline                                                
 Communication efficiency target range & $[0.82,0.86]$ \\
 Averaging period & 10 time steps\\
 Rate of change of number of cores  & 1.5 \\
 Initial number of cores & 15 \\
 Starting time step & 10 \\
 \hline
\end{tabular} \caption{Parameters for Test 3: Rate of change.} 
\label{tab:exp3} \end{table}
\begin{table}[htbp] \centering \begin{tabular}{ll}  \hline
 Parameter                        & Value   \\
 \hline                                                
 Communication efficiency target range & $[0.82,0.86]$ \\
 Averaging period & 10 time steps\\
 Rate of change of number of cores  & 3 \\
 Initial number of cores & 15 \\
 Starting time step & 10 \\
 \hline
\end{tabular} \caption{Parameters for Test 4: rate of change.} 
\label{tab:exp4} \end{table}

\begin{figure}
  \centering
  \includegraphics[width=0.49\textwidth]{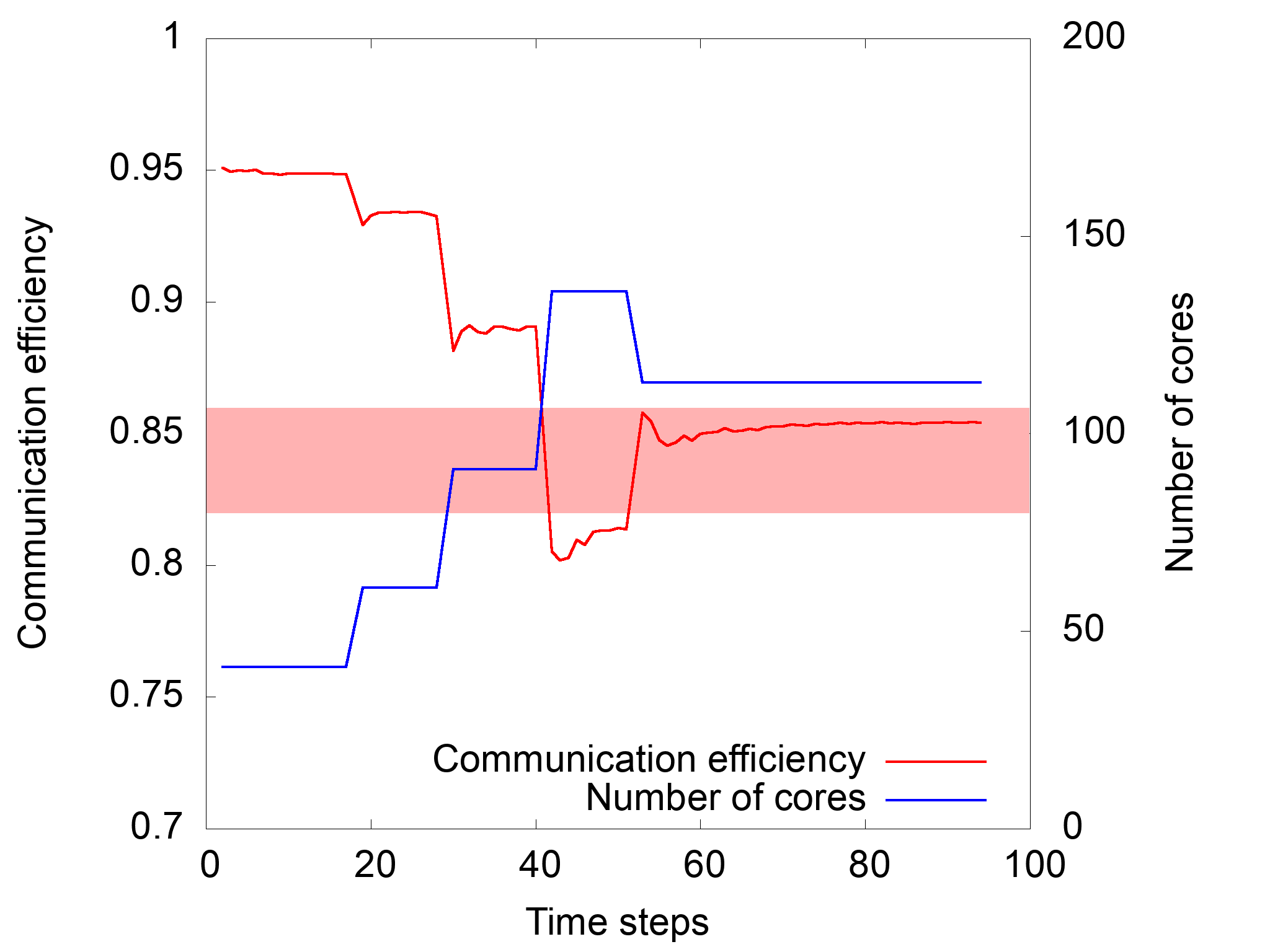}
  \includegraphics[width=0.49\textwidth]{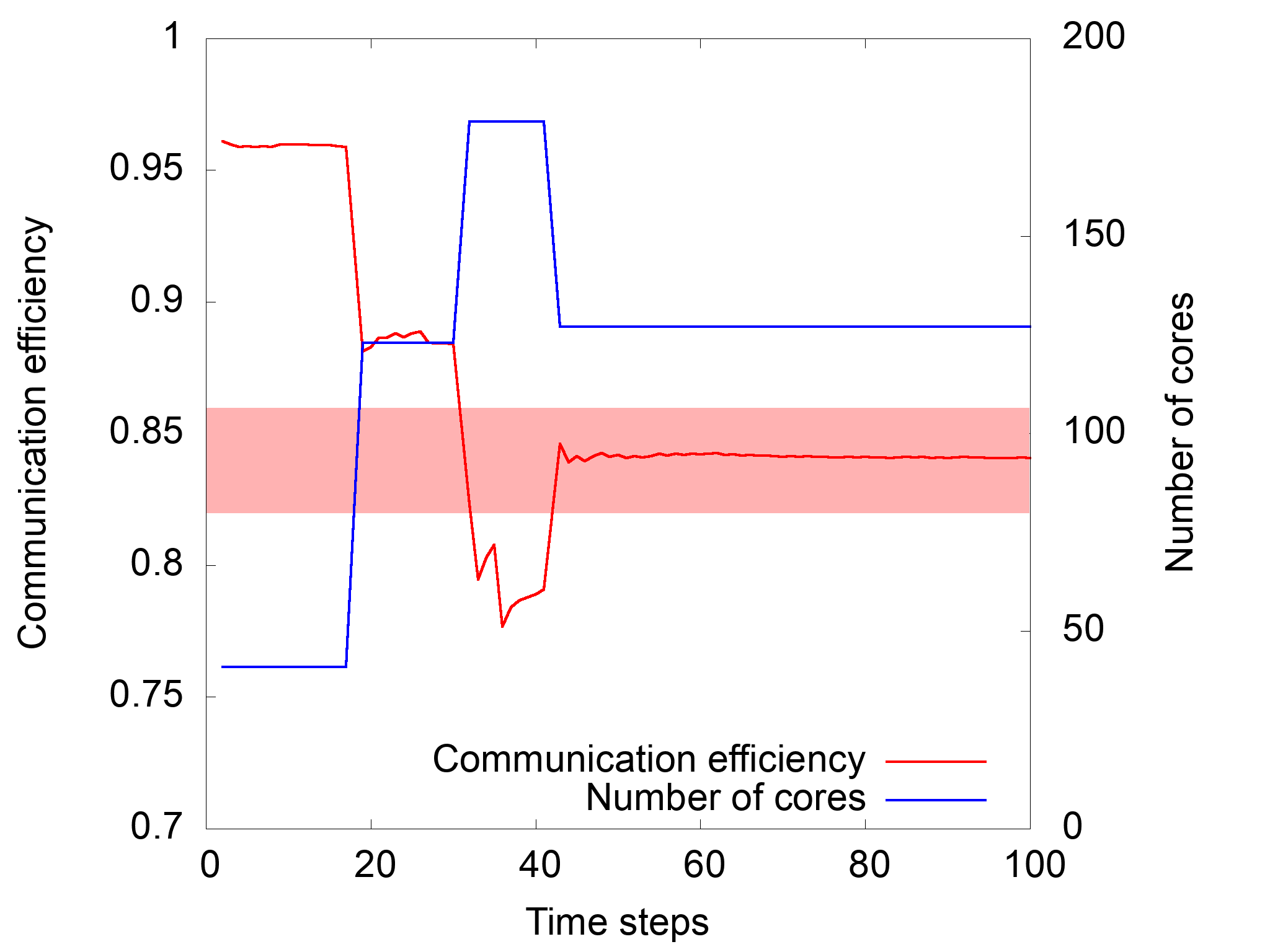}
  \caption{Optimization evolution in terms of $\CE$ and number of cores $n$ along the time integration.
  (Left)  Test 3: starting with a small rate of change (1.5). See Table \ref{tab:exp3}.
  (Right) Test 4: starting with a large rate of change (3.0). See Table \ref{tab:exp4}.
  }
  \label{fig:exp32}
\end{figure}

Figure \ref{fig:exp32} shows the evolution of the optimization strategy using a small rate of change of 1.5 and higher one 3.0, referred to as Test 3 and Test 4. For Test 3, we observe an overshoot at the third optimization step, further corrected at the fourth one. For Test 4, the target efficiency is reached in three steps, although an overshoot is also observed at optimization step 2. These tests show that the optimization is not too sensitive to the rate of change, at least for the test case considered here. 

\subsubsection{Test 5: Low target efficiency}

By lowering the target efficiency, we expect communication to have a higher weight and optimization process to be harder. The control parameters of Test 5 are given in Table \ref{tab:exp5}.
\begin{table}[htbp] \centering \begin{tabular}{ll}  \hline
 Parameter                        & Value   \\
 \hline                                                
 Communication efficiency target range & $[0.76,0.80]$ \\
 Averaging period & 10 time steps\\
 Rate of change of number of cores  & 3 \\
 Initial number of cores & 42 \\
 Starting time step & 5 \\
 \hline
\end{tabular} \caption{Parameters for Test 5: low target efficiency.} 
\label{tab:exp5} \end{table}
As in the previous example, we observe an overshoot at optimization step 2, but the target is reached at the next step. When the number of cores increases, we note more variability in the communication efficiency. Indeed, noise is more likely to affect the computation and communication when the number of cores increases, as we have a lower load per core and more probability that one of the cores is affected by noise. We finally observe that stability is eventually recovered when lowering the number of cores.
\begin{figure}
  \centering
  \includegraphics[width=0.49\textwidth]{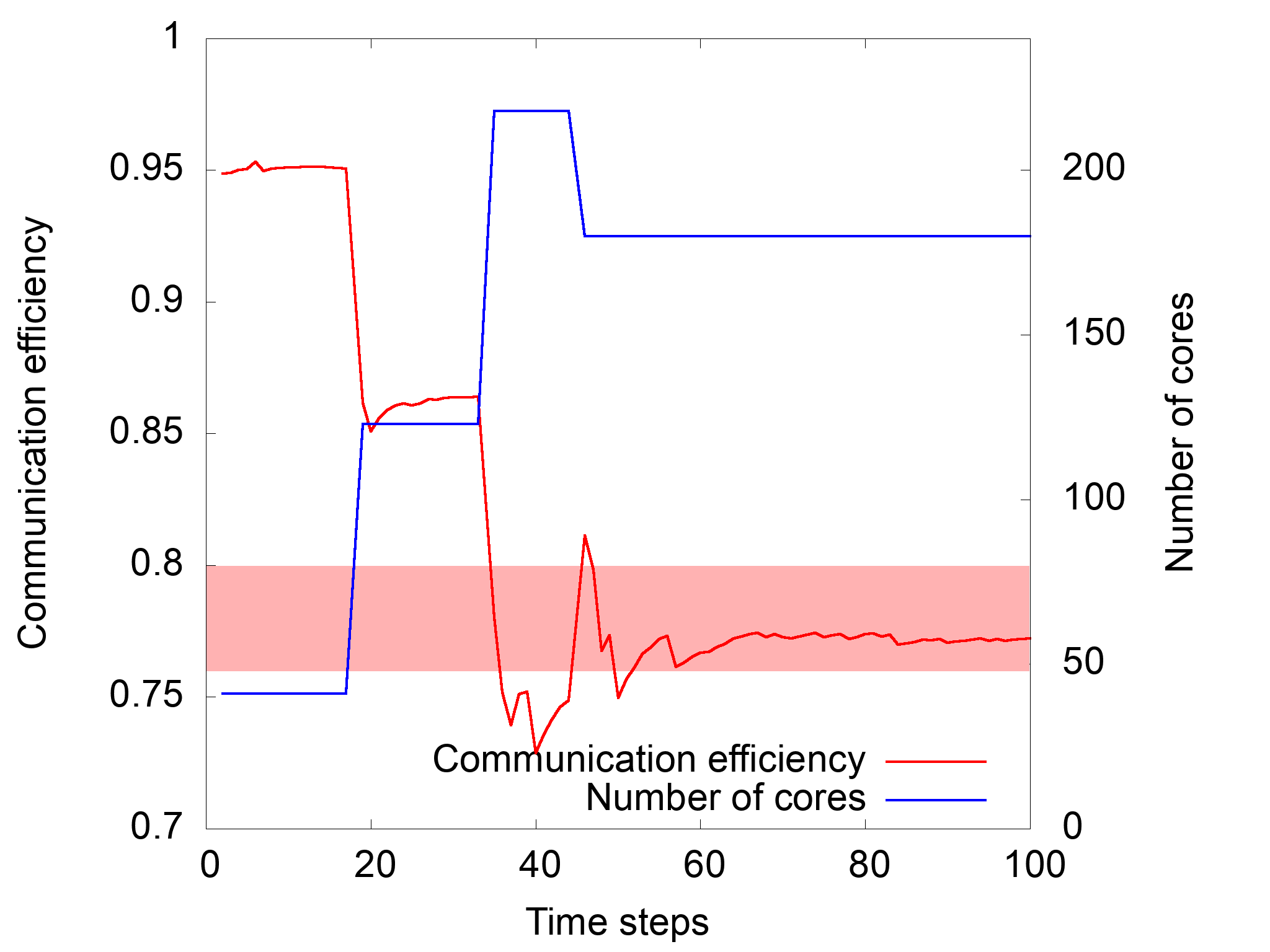}
  \caption{Optimization evolution in terms of $\CE$ and number of cores $n$ along the time integration.
  Test 5: low target efficiency. See Table \ref{tab:exp5}.
  }
  \label{fig:exp5}
\end{figure}

\subsubsection{Test 6: Varying efficiency}

\begin{figure}
  \centering
  \includegraphics[width=0.49\textwidth]{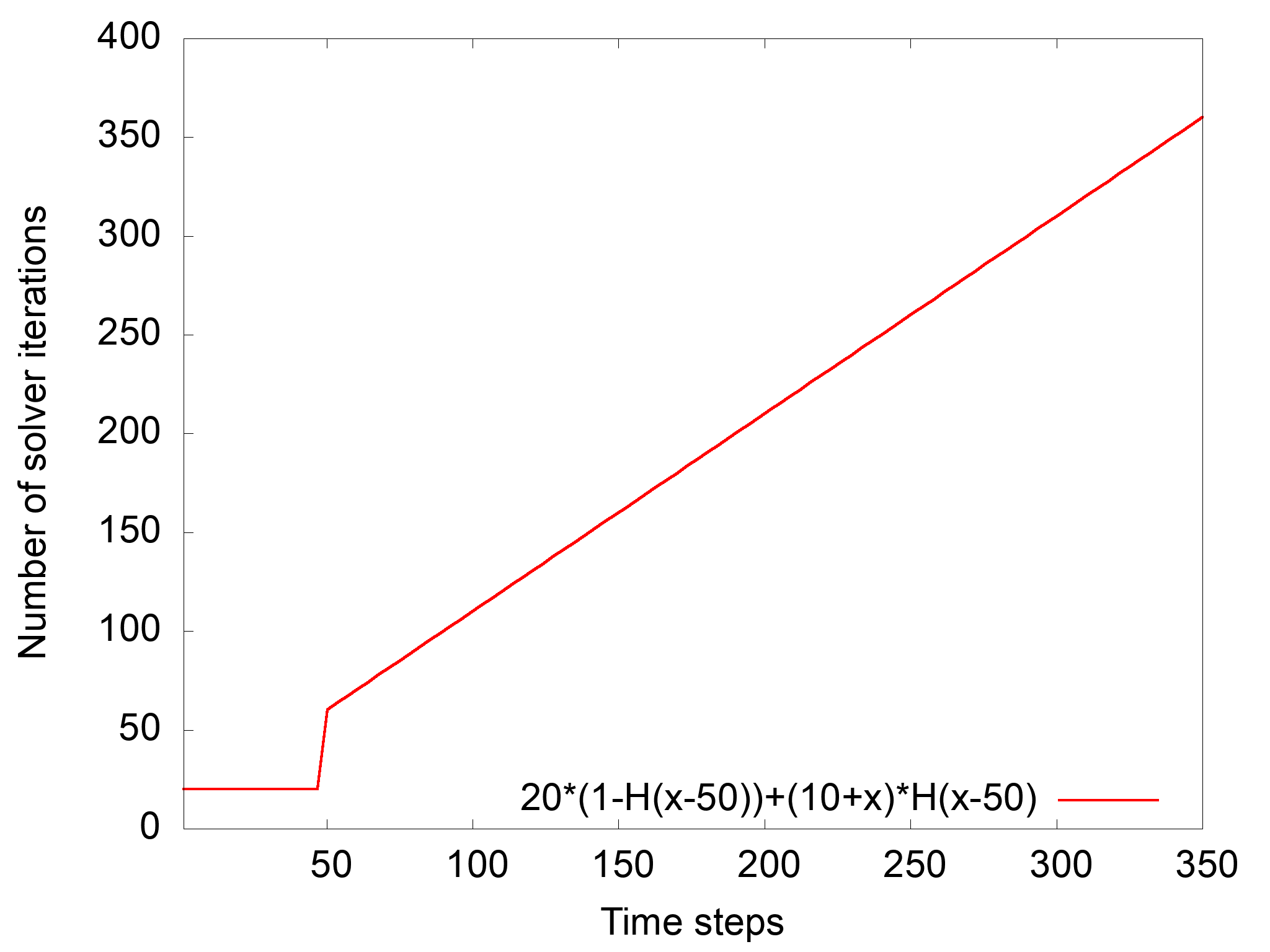}
  \includegraphics[width=0.49\textwidth]{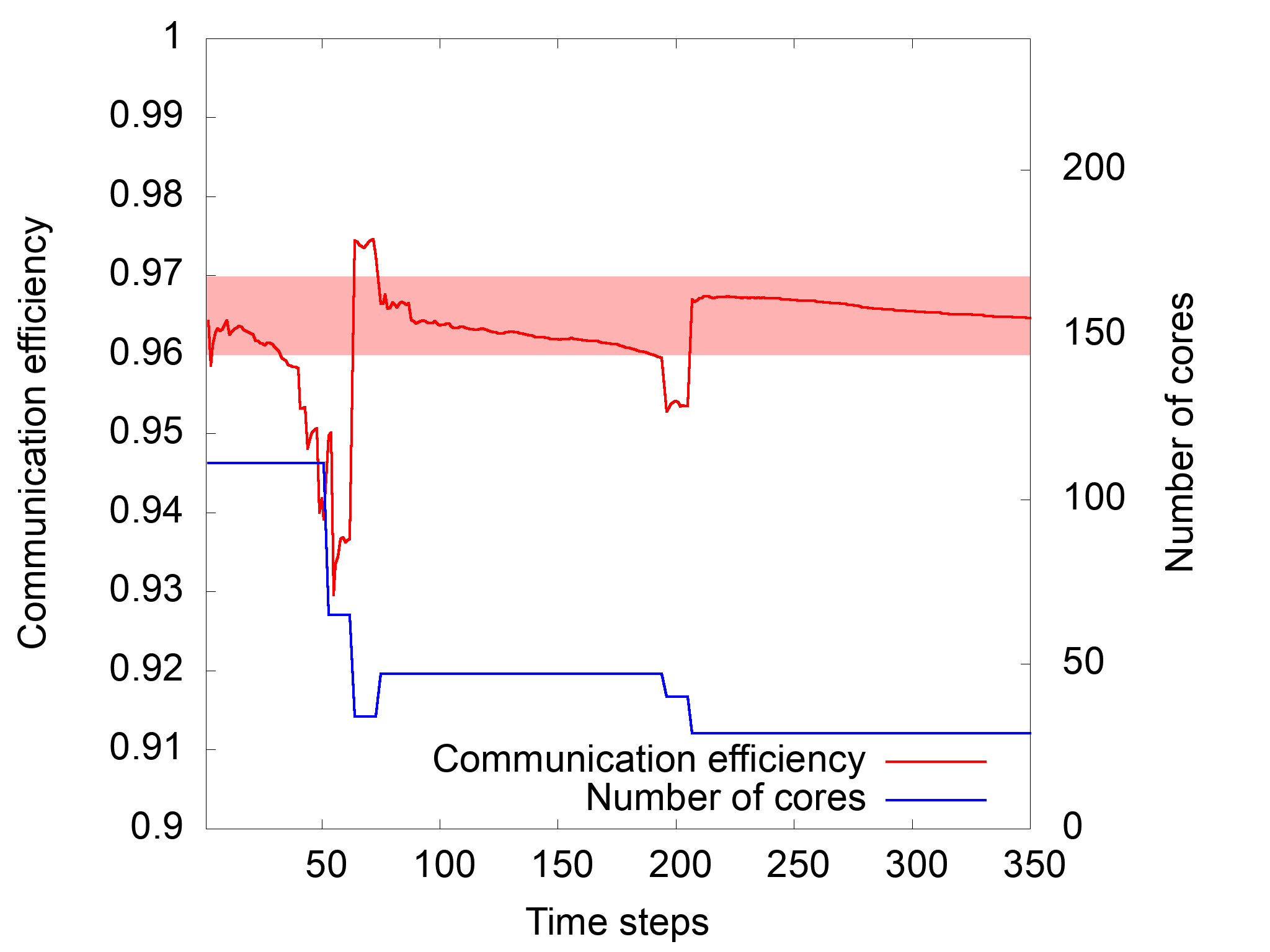}
  \caption{
  (Left)  Number of solver iterations.
  (Right) Optimization evolution in terms of $\CE$ and number of cores $n$ along with the time integration.
  Test 6: increasing number of solver iterations. See Table \ref{tab:exp6}.
  }
  \label{fig:exp6}
\end{figure}

The first five tests involved a constant communication pattern
along with the simulation because the iterations of the linear solver were almost constant. 
The linear solver is the part of the time step which contains more communication episodes, therefore the relative weight of the communications within the time-step depends on the solver iterations. In this test, we now force manually the number of iterations of the pressure solver to follow the behavior of the optimization algorithm when the communication pattern is evolving. This situation can happen in practice, when for example the time step is increased to reach a steady-state in fewer time steps, constraining the solver to carry out more iterations to converge. For this, we force the number iterations to follow this equation:
\begin{EQ}
\mbox{Number of solver iterations} = 20(1-H(x-50))+(10+x)H(x-50),
\end{EQ}
where $H(x)$ is the Heaviside function and $x$ the time step. $H(x)$ is shown in Figure \ref{fig:exp6} (Left). The control parameters used in this test are given in Table \ref{tab:exp6}.
\begin{table}[htbp] \centering \begin{tabular}{ll}  \hline
 Parameter                        & Value   \\
 \hline                                                
 Communication efficiency target range & $[0.76,0.80]$ \\
 Averaging period & 10 time steps\\
 Rate of change of number of cores  & 3 \\
 Initial number of cores & 42 \\
 Starting time step & 5 \\
 \hline
\end{tabular} \caption{Parameters for Test 6: Varying efficiency.} 
\label{tab:exp6} \end{table}

Figure \ref{fig:exp6} (Right) shows the evolution of the elastic simulation
We observe that, as the number of solver iterations increases, the communication efficiency decreases, and therefore less and fewer cores are required to maintain 
the target range. We also note that when the number of iterations of the solver changes abruptly at time step 20, the $\CE$ becomes noisy (as observed in Test 5), and the number of cores decreases very fast to accommodate for this change. However, stability is recovered quite fast.

\subsection{Test 7: Explicit scheme}

For the explicit case, communications are expected to have a relatively low weight compared to the implicit case, leading in general to better global communication efficiency. For this only test on the explicit solver, we use the control parameters given in Table \ref{tab:expl}. Note that an aggressive rate of change of 4 together with a very narrow target range has been chosen.
\begin{table}[htbp] \centering \begin{tabular}{ll}  \hline
 Parameter                        & Value   \\
 \hline                                                
 Communication efficiency target range & $[0.95,0.96]$ \\
 Averaging period & 20 time steps\\
 Rate of change of number of cores  & 4 \\
 Initial number of cores & 56 \\
 Starting time step & 10 \\
 \hline
\end{tabular} \caption{Parameters for Test 7.} 
\label{tab:expl} \end{table}

\begin{figure}
  \centering
  \includegraphics[width=0.49\textwidth]{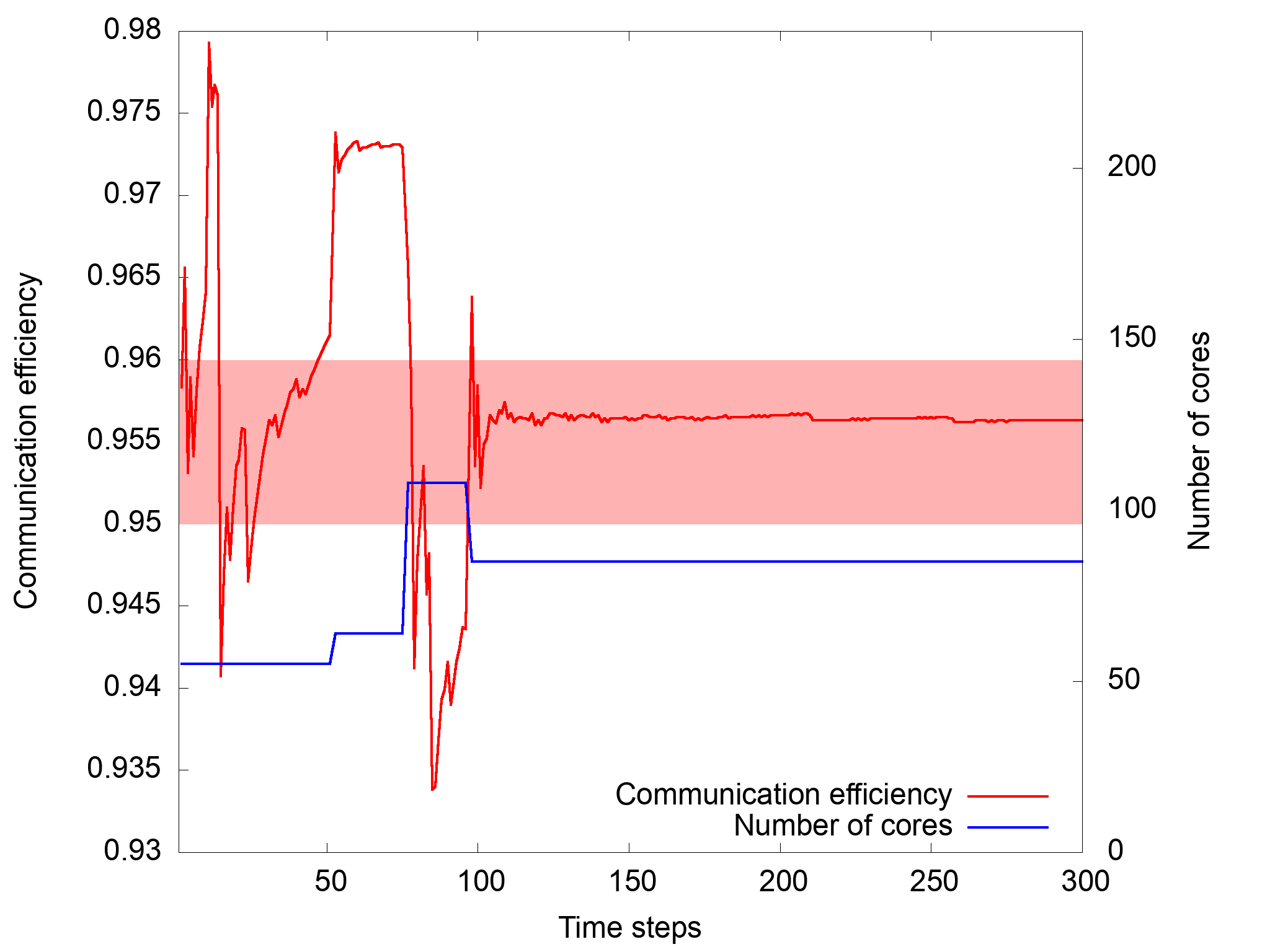}
  \caption{
  (Left)  Optimization evolution in terms of $\CE$ and number of cores $n$ along with the time integration.
  Test 7: explicit scheme. See Table \ref{tab:expl}.
  }
  \label{fig:expl}
\end{figure}

Figure \ref{fig:expl} shows a noisy communication efficiency, which should not be caused by the solver that is relatively less costly than in the implicit case.
Also, it shows an overshoot at optimization step 2, recovered just at the next step. Eventually, the simulation becomes stable after the last optimization step.

\section{Conclusions}

We have proposed a methodology to ensure a target 
communication efficiency in CFD simulations, with supporting evidence produced with the code Alya \cite{Vazquez15d}. This communication efficiency is based on real runtime measures using the library TALP. Given a target efficiency, the number of cores needed to fulfill this target is estimated at runtime. If these resources should be expanded or reduced, a workflow manager, PyCOMPSs \cite{compss_softwareX}, interacts with SLURM to provide the new amount of cores to the simulation, while remaining inside the same job. When these resources are available, the CFD 
code is restarted and the simulation continues.

The optimization workflow proposed has been validated under different situations, user requirements, and numerical methods, and has shown that the optimum partition is obtained in very few optimization steps. We would like to eventually stress that, as a side effect, the methodology proposed 
provides resilience to the simulation workflow:
if the simulation crashes for any reason, it can be resumed with the remaining available resources. 

As future work, we propose to understand the behavior of the workflow with more complex cases, where the communication efficiency patterns change during the iterations. In addition, it would be interesting to consider other target criteria, for example, the time-to-solution regardless of the communication efficiency. This option would be useful for urgent computing, attaching computing resources to the simulation as they become available. Finally, the estimate for the target number of cores could be refined, for example 
considering the measures from previous optimization steps.

\section*{Acknowledgements}

This work has been supported by the Spanish Government (Grant PID2019-107255GB-C21 by MCIN/AEI/ 10.13039/501100011033); by Generalitat de Catalunya (contract 2014-SGR-1051); by the European Commission H2020 project PoP CoE (GA n. 824080); by the European Commission H2020 project CompBioMed CoE (GA n. 823712) and by the European Commission and the EuroHPC JU under contract 955558 (eFlows4HPC project).

\bibliographystyle{elsarticle-num}
\bibliography{biblio.bib}

\end{document}